\documentclass[twocolumn]{autart} 
\newtheorem{mydef}{\bf Definition}
\newtheorem{mythm}{\bf Theorem}
\newtheorem{myprob}{\bf Problem}
\newtheorem{mylem}{\bf Lemma}
\newtheorem{mycol}{\bf Corollary}
\newtheorem{mypro}{\bf Proposition}

\newtheorem{remark}{Remark}
\newtheorem{assumption}{\bf Assumption}

\usepackage{amsfonts}
\DeclareMathSymbol{\shortminus}{\mathbin}{AMSa}{"39}
\usepackage{flushend}
\usepackage{multirow}
\usepackage{makecell}
\usepackage{booktabs}
\usepackage{array}

\usepackage{graphicx}
\graphicspath{{./Pics/}}
\DeclareGraphicsExtensions{.pdf}
\usepackage{float}
\usepackage{amsmath}
\usepackage{epsfig}
\usepackage{graphicx}
\usepackage{arydshln}
\usepackage{algpseudocode}
\usepackage{verbatim}
\usepackage{subfigure}
\usepackage{enumerate}
\usepackage{rotating}
\usepackage{enumitem}
\usepackage{color}
\usepackage{caption} 
\usepackage{mathrsfs}
\usepackage{threeparttable}
\usepackage{mathtools}
\usepackage[linesnumbered,ruled,vlined]{algorithm2e}
\usepackage{cite}
\usepackage{float}
\usepackage{soul}
\usepackage{tikz}
\usepackage{hyperref}
\hypersetup{hidelinks} 
\usepackage{soul}
\usepackage{booktabs}
\usepackage{multirow}
\usepackage{makecell}
\usepackage{array}
\usetikzlibrary{arrows.meta}
\usetikzlibrary{patterns}
\usepackage{tcolorbox}
\usepackage{fancyhdr}
\usepackage{flushend}

\allowdisplaybreaks[4]

\newfloat{tcolorboxenv}{h}{lob}

\begin{document}
\begin{frontmatter}
\title{Distributionally Robust Control Synthesis for Stochastic Systems  with Safety and Reach-Avoid Specifications}

\thanks[footnoteinfo]{This work was supported by  the National Natural Science Foundation of China (62061136004, 62173226, 61833012). Corresponding Author: Xiang Yin.}
		\author[SJTU1,SJTU2]{Yu Chen}\ead{yuchen26@sjtu.edu.cn},  
		\author[SJTU1,SJTU2]{Yuda Li}\ead{yuda.li@sjtu.edu.cn},
		\author[SJTU1,SJTU2]{Shaoyuan Li}\ead{syli@sjtu.edu.cn},
		\author[SJTU1,SJTU2]{Xiang Yin}\ead{yinxiang@sjtu.edu.cn} 
		
		\address[SJTU1]{Department of Automation, Shanghai Jiao Tong University, Shanghai 200240, China.} 
		\address[SJTU2]{Key Laboratory of System Control and Information Processing, Ministry of Education of China, Shanghai 200240, China.}

\begin{abstract}
We investigate the problem of synthesizing distributionally robust control policies for stochastic systems under safety  and reach-avoid specifications. Using a game-theoretical framework, we consider the setting where the probability distribution of the disturbance at each time step is selected from an ambiguity set defined by the Wasserstein distance. The goal is to synthesize a distributionally robust control policy that ensures the satisfaction probability exceeds a specified threshold under any distribution within the ambiguity set.
First, for both safety and reach-avoid specifications, we establish the existence of optimal policies by leveraging the
dynamic programming principles. Then we demonstrate how the associated optimization problem can be efficiently solved using the dual representation of Wasserstein distributionally robust optimization. Furthermore, for safety specifications in particular, we introduce a novel concept of distributionally robust control barrier certificates and show how these enable the efficient synthesis of controllers through sum-of-squares programming techniques. 
Finally, our experimental results reveal that incorporating distributional robustness during the synthesis phase significantly improves the satisfaction probability during online execution, even with limited statistical knowledge of the disturbance distribution.
\end{abstract}
\end{frontmatter}

\section{Introduction} 
\subsection{Motivations and Backgrounds}

In recent years, formal controller synthesis for  cyber-physical systems has garnered significant attention due to its fundamental role in ensuring safety for critical applications such as autonomous vehicles, robotic systems, and manufacturing systems~\cite{belta2019formal,pola2019control,yin2024formal}.
A common feature of these systems is their operation in open environments that are subject to disturbances or even adversarial inputs. For instance, a navigation robot must ensure it reaches its target region while avoiding obstacles, regardless of uncertainties encountered during operation \cite{kress2018synthesis}. 
Ensuring the satisfaction of desired tasks with provable correctness guarantees is a challenging problem  due to the complex dynamics of these systems and the unpredictability of their operating environments. 

To ensure the satisfaction of design objectives in uncertain environments, existing works on formal controller synthesis can generally be categorized into two main frameworks:\vspace{-6pt}
\begin{itemize}[leftmargin=*,topsep=0pt, itemsep=10pt]
\item \emph{Robust Control Framework: }
In this setting, it is assumed that environment disturbances lie within a bounded set without additional information. The control problem is formulated to address the worst-case scenario, ensuring that the system can achieve its design objective regardless of the disturbances encountered \cite{kress2018synthesis,cassandras2021introduction}. This approach essentially corresponds to a zero-sum game setting, where the disturbance acts as an adversary against the control player. While this method provides strong guarantees for the worst-case scenarios, it is often overly conservative because it does not leverage any prior knowledge about the disturbances. 
    \item
\emph{Stochastic Control Framework: }
In this setting, environmental disturbances are modeled as random variables with known statistical characteristics. The control objective is typically to maximize the probability of achieving the task or to ensure that the satisfaction probability exceeds a given threshold \cite{bertsekas1996stochastic,cao2008stochastic}. However, this approach faces practical challenges, such as the difficulty or high cost of obtaining accurate disturbance distributions. Additionally, it may lack robustness, as the actual underlying environment may deviate from the assumed nominal distribution.\vspace{-6pt}
\end{itemize}

More recently, benefiting from advancements in distributionally robust (DR) optimization techniques  from the operational research community, the distributionally robust control framework has been proposed to bridge the gap between robust control and stochastic control frameworks \cite{delage2010distributionally,xu2012distributionally,van2015distributionally}. In the DR setting, environmental disturbances are still modeled as random variables, but their distributions are not fully known a priori. Instead, it is assumed that the disturbance distribution lies within an ambiguity set, and the objective is to optimize the worst-case expected performance over this set.
The advantage of this framework lies in its ability to leverage statistical information about the environment while maintaining robustness against distributional shifts or knowledge mismatches. Moreover, it facilitates a data-driven approach to control synthesis, as the statistical information required for the ambiguity set is often derived from empirical data.

\subsection{Our Results and Contributions}
In this paper, we address the formal control synthesis problem for stochastic systems with safety and reach-avoid specifications. Specifically, we consider discrete-time stochastic control systems and adopt the distributionally robust framework by assuming that the actual disturbance is a random variable whose distribution lies within an ambiguity set characterized by the Wasserstein distance from a nominal distribution. The control objective is to ensure the robust satisfaction of the safety or reach-avoid task, guaranteeing that the satisfaction probability always exceeds a given threshold, regardless of the actual distribution within the ambiguity set. 

We adopt a game theoretical  approach by formulating the control problem as a two-player game. At each time step, the controller selects a control input, while the environment player chooses a distribution from the ambiguity set. This dynamic game formulation precisely captures scenarios where the environment distribution is time-varying but bounded, and provides a conservative yet computationally feasible solution when the environment distribution is unknown but fixed.  Our technical results and contributions are summarized as follows:\vspace{-3pt}
\begin{itemize}[leftmargin=*,topsep=0pt, itemsep=10pt]
\item 
First, for both safety and reach-avoid specifications, we prove the existence of optimal control policies under Wasserstein ambiguity sets by characterizing the optimality conditions through dynamic programming equations. This result extends existing works on reach-avoid dynamic games beyond the purely stochastic setting and generalizes distributionally robust dynamic games from safety specifications to the more complex reach-avoid specifications.
\item 
We then provide a computationally feasible approach for solving the optimization problems arising in the dynamic programming equations. Our approach is based on a duality reformulation, which effectively reduces the original infinite-dimensional optimization problems to  finite-dimensional ones by leveraging the finite support of the nominal distribution.
\item 
Finally, for safety specifications, we further propose a computationally more efficient approach to solve the control synthesis problem. 
This approach utilizes the new notion of distributionally robust control barrier certificates to approximate a lower bound on the satisfaction probability. By doing so, the dual optimization problems can be further transformed into sum-of-squares  programs, which are more computationally tractable and can be solved efficiently.
\end{itemize}

\subsection{Related Works}

\emph{Formal Controller Synthesis: }
There has been significant recent progress in controller synthesis for formal specifications with performance guarantees. A common approach is the abstraction-based method, which constructs finite abstractions of concrete systems and then employs symbolic and algorithmic techniques for controller synthesis; see, e.g., ~\cite{tabuada2009verification, zamani2014symbolic, belta2017formal}. The key advantage of this approach is its ability to algorithmically handle complex formal specifications, such as temporal logic. However, a main challenge lies in its computational scalability, particularly for high-dimensional systems. To address this computational challenge, abstraction-free methods such as control barrier certificates have gained traction. Control barrier certificates provide sufficient yet computationally efficient conditions to ensure formal specifications. This approach has been applied to various classes of systems, including continuous-time systems~\cite{ames2019control, santoyo2021barrier, meng2024stochastic} and discrete-time systems~\cite{jagtap2020formal, santoyo2021barrier, chen2023data}.
In our work, we also build upon this general idea to address the safe control synthesis problem more efficiently. However, to the best of our knowledge, control barrier certificates have not yet been applied to the distributionally robust control setting.  

\emph{Stochastic Control and Games: }
To precisely analyze the solvability of formal control synthesis problems for hybrid dynamic systems under disturbances, one needs, in principle, to solve the corresponding Hamilton-Jacobi (HJ) equations~\cite{bansal2017hamilton}. Specifically, the feasibility of the specification under disturbances can be characterized as a level set of the PDE solution. For example, safety tasks are addressed in~\cite{mitchell2005time}, while reach-avoid tasks are investigated in~\cite{fisac2015reach}. In general, the PDE solutions can be computed numerically~\cite{mitchell2008flexible} or approximated using neural networks~\cite{bansal2021deepreach}.
For stochastic discrete-time systems, dynamic programming~\cite{bertsekas1996stochastic} is commonly used to compute the probabilities for safety~\cite{abate2008probabilistic} and reach-avoid~\cite{summers2010verification} specifications. 
Notably, in~\cite{ding2013stochastic}, the authors propose a general stochastic dynamic game framework to characterize and efficiently solve safety and reachability controller synthesis problems with probabilistic guarantees.
Our work is also motivated by these existing approaches. However, these methods typically assume that the disturbance follows a known distribution. In contrast, we address the distributionally robust setting, which is more general. This generalization is non-trivial, as it requires establishing the measurability of policies and proving the existence of optimal solutions under different information structures.  

\emph{Distributionally Robust Control Synthesis: }
Finally, there is growing interest in distributionally robust formal verification and synthesis; see, e.g.,~\cite{yang2018dynamic, romao2023distributionally, schon2024data, gracia2025efficient}.
For example, \cite{romao2023distributionally} tackles the DR verification problem using kernel conditional mean embedding, while our work focuses on DR control synthesis.
In the context of control synthesis, \cite{gracia2025efficient} study the Wasserstein DR control synthesis problem for switched stochastic systems under temporal logic specifications. However, their approach is limited to finite control inputs, where the controller switches the system between a finite number of modes. The stochastic system model considered in our work is more general, and our methods differ significantly from their abstraction-based approaches.
Our work is also closely related to~\cite{yang2018dynamic}, which addresses the DR control synthesis problem for safety specifications. While \cite{yang2018dynamic} focuses on moment uncertainty, we consider Wasserstein uncertainty. Additionally, we extend the framework to include reach-avoid specifications and require fewer assumptions for safety specifications. Moreover, we also propose  barrier certificate-based synthesis approach, which is not investigated in \cite{yang2018dynamic}.

\subsection{Organizations}
The rest of the paper is organized as follows. 
Section~\ref{sec:prelinimary} provides the basic preliminaries, and Section~\ref{sec:formulation} introduces the problem formulation. Section~\ref{sec:dynamicprogramming} presents the general solution for reach-avoid and safety specifications using dynamic programming. 
In Section~\ref{sec:approximate}, we focus on the safety case and introduce distributionally robust control barrier certificates as a sufficient yet efficient tool for synthesizing controllers. 
Finally, we illustrate the proposed methods by two case studies in Section~\ref{sec:casestudy} and conclude the work in Section~\ref{sec:con}.

\section{Preliminary}\label{sec:prelinimary}
\textbf{Notations: }
We denote by $\mathbb{R}$, $\mathbb{R}_{\geq 0}$ the set of all real numbers and non-negative real numbers, respectively. Given a Borel space $A$, $\mathcal{B}(A)$ and $\mathcal{P}(A)$ represent its Borel $\sigma$-algebra
and the set of Borel probability measures on $A$, respectively.  
Unless otherwise stated, all sets of probability measures are endowed with the weak topology \cite[Sec. 7.4.2]{bertsekas1996stochastic}.  Let $Y$ and $Z$ be separable and metrizable spaces. A stochastic kernel $q(\mathrm{d}z|y)$ on $Z$ given $y$ is a collection of probability measures in $\mathcal{P}(Z)$ parameterized by $y \in Y$. Given $B \in \mathcal{B}(Z)$, we denote by $q(B|y)$ the probability on $B$ under measure $q(\mathrm{d}z|y)$. For simplicity we write Borel measurable directly as measurable.

\subsection{Discrete-Time  Stochastic Control Systems}
In this work, we consider a discrete-time  stochastic control system, which is a 6-tuple 
    \begin{equation}\label{eq:system}
        \mathbb{S}=(X,X_0,U,W,f,T),
    \end{equation}
    where\vspace{-6pt}
    \begin{itemize}
    \item 
       $X=\mathbb{R}^n$ is the state space;\vspace{3pt}
    \item  
       $X_0 \subseteq X$ is the set of initial states;\vspace{3pt}
    \item 
       $U \subseteq \mathbb{R}^m$ is a compact  set of all control inputs;\vspace{3pt}
    \item 
        $W \subseteq \mathbb{R}^l$ is a compact set  representing the support of disturbances;\vspace{3pt}
    \item 
        $f:X \times U \times W \to X$ is continuous function  representing the dynamic of the system;\vspace{3pt}
    \item 
        $T$ is a positive integer representing the horizon of the system.
    \end{itemize} 
In this work, we consider an adversarial decision-making setting, i.e., the disturbance $w_t \in W$ at each time instant $t$ is controlled by an adversarial environment. However, instead of considering the non-stochastic game setting, where the environment can decide the disturbance directly, we assume that the environment can only decide the distribution $\mu_t \in \mathcal{P}(W)$ of disturbance $w_t \in W$.  
We assume that the controller has prior information that the distribution of the disturbance belongs to an ambiguity set $\mathbb{D} \subseteq \mathcal{P}(W)$ but cannot determine the actual distribution at each time instant.

Formally, a \emph{control policy} $\pi$ is a sequence  
\begin{equation}
\pi=(\pi_0,\pi_1,\dots,\pi_{T-1})
\end{equation}
such that each $\pi_t: X \to U$ is measurable.
Although the policy should generally be universally measurable to preserve measurability in dynamic programming~\cite[Page 156]{bertsekas1996stochastic}, a Borel measurable policy is sufficient in our work since the value functions considered are always upper semi-continuous.
We denote by $\Pi$ the set of control policies. 
Similarly, an \emph{adversary  strategy} $\gamma$ is a sequence
\begin{equation}
\gamma=(\gamma_0,\gamma_1,\dots,\gamma_{T-1})
\end{equation}
such that each   $\gamma_t: X \times U \to \mathbb{D}$ is measurable. 
That is, it chooses the distribution of disturbances based on the current information. We denote by $\Gamma$ the set of adversary strategies. Note that, since $\mathbb{D}$ is still a topological space with the relative topology induced by $\mathcal{P}(W)$, the measurability of adversary strategy is well-defined. 
Moreover, we only consider Markov policy because from~\cite{rieder1991non}, when the transition probability is Markov and the utility function is sum-multiplicative (as in our case), it is sufficient to consider the class of Markov policies.

Given control policy $\pi \in \Pi$, adversary strategy $\gamma \in \Gamma$,
the induced state stochastic kernel at time $i$ is
$\mathbf{k}_{i}: X  \to \mathcal{P}(X)$ such that, for any  $ B  \in \mathcal{B}(X)$, we have
\begin{equation}\label{eq:state stochastic kernel}
\mathbf{k}_{i}(B\mid x_{i})\!=\! 
 \gamma_{i}(x_{i},\pi_{i}(x_{i}))(\{ w \!\in\! W \mid f(x_{i},\pi_{i}(x_{i}),w) \!\in\! B\}).
\end{equation}
Without loss of generality, we do not consider the distribution of the initial state further. Instead, our objective is to design a control policy that satisfies certain specifications for all possible initial states. 
To this end, for each initial state $x_0\in X_0$, we denote by $\textsf{Pr}_{x_0}^{\pi,\gamma} \in \mathcal{P}(\{x_0\} \times X_1 \times X_2 \cdots \times X_T)$ the unique probability measure induced by the stochastic kernels $\mathbf{k}_0, \dots, \mathbf{k}_{T-1}$ over the sample space of all state trajectories starting from $x_0$ with horizon $T$. The existence of probability measure $\textsf{Pr}_{x_0}^{\pi,\gamma}$ is proved by Lemma~\ref{lemma:measurable} in the Appendix~\ref{appendix:measurable}.

\subsection{Distributionally Robust Formal Specifications}
In this work, we consider two types of formal specifications:
(i) the \emph{reach-avoid} specification, which requires the system to reach a target set while remaining within a safety set; and
(ii) the \emph{safety} specification, which requires the system to stay  within a safety set during the entire horizon. 
The formal definitions are as follows.

\begin{mydef} 
Let $S \in \mathcal{B}(X)$ be a  Borel set representing  the safe region
and  $G \in \mathcal{B}(X)$ be a  Borel set representing  the target region   with $G \subseteq S$. 
Let  $\pi$ be a  control policy,  $\gamma$ be an adversary  strategy, and $x_{0}\in X_0$ be an initial state. 
We denote by $\mathbf{x}_{0:T}=(x_0,x_1,\dots,x_T)$ the state trajectory from $x_0$.  
The probability of satisfying the safety specification (or safe probability) is defined by
\begin{equation} \label{eq:safetydef}
\textsf{SA}_{x_0}^{\pi,\gamma}(S):=\textsf{Pr}_{x_0}^{\pi,\gamma}(\{ \mathbf{x}_{0:T} \mid
           \forall t \in [0,T], x_{t} \in S \}), 
\end{equation}   
and the probability of  satisfying  the reach-avoid specification  (or reach-avoid probability) is defined by
\begin{align} \label{eq:reach-avoiddef} 
&\textsf{RA}_{x_0}^{\pi,\gamma}(G,S):=\\
&\textsf{Pr}_{x_0}^{\pi,\gamma}(\{  \mathbf{x}_{0:T}  \mid 
\exists t \!\in\! [0,T],   (x_t \!\in\! G  \!\wedge\!  (\forall t' \!\in\! [0,t], x_{t'} \!\in\! S)) \}).\nonumber
\end{align} 
\end{mydef}
Note that, the above defined safety and reach-avoid probabilities can be reformulated \cite{abate2008probabilistic,summers2010verification,ding2013stochastic}  as 
\begin{align} 
 \textsf{SA}_{x_0}^{\pi,\gamma}(S) & := 
\mathbb{E}^{\pi,\gamma}_{x_0}\left[   \prod_{i=0}^{T} \mathbf{1}_{S}(x_i)  \right], \nonumber \\
    \textsf{RA}_{x_0}^{\pi,\gamma}(G,S)&:= \label{eq:reach-avoidreformulate} \\
\mathbb{E}^{\pi,\gamma}_{x_0} &\left[ \mathbf{1}_{G} (x_0)+\sum_{j=1}^T \left( \prod_{i=0}^{j-1} \mathbf{1}_{S\setminus G}(x_i)\right) \mathbf{1}_G(x_j)  \right], \label{eq:safetyreformulate}
\end{align}
respectively, where $\mathbb{E}^{\pi,\gamma}_{x_0}$ denotes the expectation with respect to the probability measure $\textsf{Pr}_{x_0}^{\pi,\gamma}$ and $\mathbf{1}_{X'}:X\to \{0,1 \}$ is the indicator function such that $\mathbf{1}_{X'}(x)=1$ if and only if $x\in X'$.

Finally, since the controller only knows that the action space of the adversary is an ambiguity set, 
we need to consider the worst-case probability over all possible adversary  strategies.  

\begin{mydef} 
Given initial state $x_0 \in X_0$ and control policy $\pi \in \Pi$, the \emph{worst-case safe probability} is defined as
\begin{equation}\label{eq:worstcasereachavoid}
    \textsf{SA}_{x_0}^{\pi}(S):=\inf_{\gamma \in \Gamma}\textsf{SA}_{x_0}^{\pi,\gamma}(S).
\end{equation}
The \emph{optimal} safety probability is defined by
\begin{equation}\label{eq:optimalreachavoidpro}
\textsf{SA}^\star_{x_0}(S):= \sup_{\pi\in \Pi} \textsf{SA}_{x_0}^{\pi}(S).
\end{equation}
Similarly, we define the worst-case reach-avoid probability for $\pi \in \Pi$ and the optimal reach-avoid probability as $\textsf{RA}_{x_0}^{\pi}(G,S)$ and $\textsf{RA}^\star_{x_0}(G,S)$, respectively.
\end{mydef}

\subsection{Wasserstein Ambiguity Sets}
In principle, there are many different ways to define the ambiguity set $\mathbb{D} \subseteq \mathcal{P}(W)$ for the action space of the adversarial environment. In this work, we adopt the \emph{Wasserstein metric}, which is widely used to characterize the distance between probability distributions, to construct the ambiguity set.
Formally, we assume that there is a nominal distribution $\nu_{N}\in \mathcal{P}(W)$, and the actual distribution of the disturbance $w_t\in W$ at each time instant lies within the Wasserstein ball centered at $\nu_{N}$ with radius $\theta > 0$:
\begin{equation}\label{eq:ambiguitydef}
    \mathbb{D}:=\{ \mu \in \mathcal{P}(W)\mid \mathcal{W}_p(\mu,\nu_{N}) \leq \theta \}.
\end{equation}
Here, $\mathcal{W}_p(\mu,\nu_{N})$ is the Wasserstein metric
of order $p \in [1,\infty)$ such that
\begin{equation}\label{eq:wasserdef}
    \begin{aligned}
&\mathcal{W}_p(\mu,\nu_{N}):=\\
&\min_{\kappa \in \mathcal{P}(W^2)}\Bigg\{ \Bigg[ \int_{W^2}  d^p(w,w') \kappa(\mathrm{d} w,\mathrm{d}w') \Bigg]^{\frac{1}{p}}   
\!\! \mid \!\!
 \begin{array}{ll}
      & \mathbb{M}^1(\kappa)=\mu   \\
      & \mathbb{M}^2(\kappa)=\nu_{N}
 \end{array}\!\! \Bigg\}
    \end{aligned}
\end{equation}
where $d$ is a metric on $W$ and $\mathbb{M}^i(\kappa)$ denotes the $i$-th marginal of $\kappa$ for $i=1,2$.
Intuitively, the Wasserstein distance between two probability distributions represents the minimum cost of transporting mass from one to the other via non-uniform perturbation, and the optimization variable $\kappa$ can be interpreted as a transport plan \cite{gao2023distributionally}.
 The Wasserstein distance offers several advantages, such as leveraging more information from the nominal distribution than the moment-based approach, and encompassing a wider range of relevant distributions compared to many $\phi$-divergences, such as Kullback-Leibler divergence. For further details on the advantages of the Wasserstein ambiguity set, readers are referred to~\cite{gao2023distributionally}.

\section{Problem Formulation}\label{sec:formulation}
In this work, our objective is to find a control policy such that the system can achieve the specification with probability higher than a threshold from any initial state. 
The problem formulations are stated as follows.
\begin{myprob}\label{problem:threshold}
Consider discrete-time stochastic control system $\mathbb{S}=(X,X_0,U,W,f,T)$,
probability threshold $\alpha \in [0,1]$, 
Wasserstein ambiguity set $\mathbb{D}$ in \eqref{eq:wasserdef} centered at $\nu_{N}$ with radius $\theta > 0$. 
Find control policy $\pi^\star \in \Pi$ such that
$\mathsf{SA}_{x_0}^{\pi^\star}(S)\geq \alpha$ for all $x_0 \in X_0$.
\end{myprob}
We denote the problem formulated above by Problem~\ref{problem:threshold}-($\textsf{SA}$). 
If $\textsf{SA}_{x_0}^{\pi^\star}(S)$ is replaced by $\textsf{RA}_{x_0}^{\pi^\star}(G,S)$, 
then the problem considers the reach-avoid specification and is denoted by Problem~\ref{problem:threshold}-($\textsf{RA}$).

Finally, we explain the physical meaning of the problem formulation, as well as the nominal distribution $\nu_N$. 
In practice, the proposed distributionally robust stochastic game setting can capture the following two scenarios:\vspace{-6pt}
\begin{itemize}[leftmargin=*,topsep=0pt, itemsep=10pt]
\item 
\emph{Purely Antagonistic Adversary:}
The adversary player is purely antagonistic and can take varying actions to change the stochastic kernel within the ambiguity set $\mathbb{D}$. 
In this setting, the nominal distribution $\nu_N$ essentially represents the center of the ambiguity set, and the radius $\theta$ captures the adversary's potential to alter the distribution.\vspace{-3pt}
\item\emph{Unknown Disturbance: }
In some applications, the decision of the adversary player is unknown but fixed over time. 
In this case, we can collect empirical data from historical executions to estimate the underlying unknown distribution. In such scenario,  $\nu_N$ is typically an empirical distribution, and the radius $\theta$ accounts for robustness against the uncertainty in the empirical distribution. In this setting, our problem can be interpreted as a conservative approach for the actual scenario, where the distribution does not change, but we consider it within a dynamic game context. 
\end{itemize}
One may regard that the adversary player should use a stationary policy when it is actually an unknown disturbance. However, using a time-invariant adversary setting will lead to an intractable program since the original decision problem involves multi-step dynamic system evolution with decision variables at each step. This setting intends to compress the complex multi-stage decision problem into a high-dimensional single-stage problem. The high-dimensional issue must be addressed to adopt a less conservative setting.

Hereafter, throughout the paper, we further assume that the nominal distribution $\nu_N$ is constructed based on a finite support.
Formally, let $\hat{w}_1, \dots, \hat{w}_M \in W$ be the supports of the nominal distribution, and for each $i = 1, \dots, M$, let $p_i \in (0,1)$ be the corresponding probability, such that $\sum_{i=1}^M p_i = 1$. 
\begin{assumption} 
The nominal distribution takes the following form:
\begin{equation}\label{eq:finitesupportdistribution}
    \nu_N = \sum_{i=1}^M p_i \delta_{\hat{w}_{i}},
\end{equation}
where $\delta_{\hat{w}_{i}}$ is  the Dirac delta measure concentrated at $\hat{w}_{i}$.
\end{assumption}
As discussed above, in the case of an unknown disturbance, the nominal distribution typically comes from the empirical data. 
In the case of an antagonistic adversary, $\nu_N$ is a user-defined distribution based on prior knowledge and confidence regarding the adversary. 
Note that even though the nominal distribution is finite-support, the Wasserstein ambiguity set still contain continuous distribution since the Wasserstein distance is possible to be finite between discrete and continuous distributions.

\section{General Solutions by Dynamic Programming}\label{sec:dynamicprogramming}
In this section, we investigate the solvability of   Problem~\ref{problem:threshold}. 
First, we prove the   existence of optimal control policy for reach-avoid specifications by dynamic programming. Then  we show how   the optimal policy can be computed effectively  using the dual method of Wasserstein distributionally robust optimization.
Finally, we extend our results to the case of safety specifications.

\subsection{Existence of Optimal Control Policies}\label{subsection:markov}
First, we focus on the reach-avoid specification. 
In order to show the existence of the optimal control policy, we define a dynamic programming operator as follows.   Let $\mathbf{v}$ be a measurable function on $X$.
We define operator $\mathbb{T}$ by: for any $x\in X$, we have
\begin{equation}\label{eq:operator-1}
\mathbb{T}(\mathbf{v})(x):=\mathbf{1}_{G}(x) + \mathbf{1}_{S\setminus G}(x) \sup_{u \in U} \inf_{\mu 
    \in \mathbb{D}} \mathbf{H}(x,u,\mu,\mathbf{v}), 
\end{equation}
where \vspace{-6pt}
\begin{equation}\label{eq:operator-2}
 \mathbf{H}(x,u,\mu,\mathbf{v})= \int_{W} \mathbf{v}(f(x,u,w)) \mathrm{d} \mu(w).
\end{equation}
Based on the above defined operator $\mathbb{T}$, we define a sequence of value functions  
$\{{\mathbf{v}}_t\}_{t=0,1,\dots,T}$ by:
\begin{align} \label{eq:valuefunctionateachstage}
		\left\{
		\begin{array}{ll }
  \mathbf{v}_T=\mathbf{1}_G\\
			 \mathbf{v}_t=\mathbb{T}(\mathbf{v}_{t+1}), \forall t=0,1,\dots, T-1
		\end{array}. 
		\right.   
\end{align}
Intuitively, $\mathbf{v}_0(x)$ characterizes the reach-avoid probability starting from state $x\in X$ under optimal control policy and adversary strategy.
First, we show that the integration in Eq.~\eqref{eq:operator-2} is well-defined when sets $G$ and $S$ are compact. 
\begin{mypro}\label{prop:uppersemi}
Assume that sets $G$ and $S$ are closed. 
Then for each $t=0,1,\dots, T$, function $\mathbf{v}_t$ is upper semi-continuous, which means that $\mathbf{H}(x,u,\mu,\mathbf{v}_t)$ is well-defined. 
Furthermore, we have $\mathbf{v}_t(x) \in [0,1]$ for $x \in X$.\vspace{-6pt}
\end{mypro}
\begin{pf}
    The proof is provided in the Appendix~\ref{appdendix:mainbodyproof}. 
\end{pf} 

Hence, hereafter we make the following assumption. 
\begin{assumption}\label{assumption:compact}
   The sets $G$ and $S$ in \eqref{eq:reach-avoiddef} are blue.
\end{assumption}

Next, we show that in Eq.~\eqref{eq:operator-1}, the controller can exactly attain the supremum, while the adversary can only adopt a strategy that is sub-optimal. 
\begin{mypro} \label{prop:optimalcontrolandsuboptimaladv}
For each stage  $t=0,1,\dots, T-1$ and the value function $\mathbf{v}_t$ defined in  Eq.~\eqref{eq:valuefunctionateachstage},
    \begin{enumerate} 
    \item 
    There is measurable function 
    $\pi^\star_t:X\to U$ such that, for any $ x \in X$, we have 
    \begin{equation}\label{eq:max-achieve}
    \pi^\star_t(x) \in \arg \max_{u \in U} \inf_{\mu \in \mathbb{D}} \mathbf{H}(x,u,\mu,\mathbf{v}_{t+1}).     
    \end{equation} 
    \item 
    For any $\epsilon >0$, there is measurable function 
    $\gamma^\star_t: X\!\times\! U \!\to\! \mathbb{D}$ such that, for any $(x,u) \in X \!\times\! U$, we have 
    \[
    \mathbf{H}(x,u,\gamma^\star_t(x,u),\mathbf{v}_{t+1}) \leq \inf_{\mu \in \mathbb{D}} \mathbf{H}(x,u,\mu,\mathbf{v}_{t+1})+\epsilon. 
    \] 
\end{enumerate}
\end{mypro}
\begin{pf}
    The proof is provided in the Appendix~\ref{appdendix:mainbodyproof}. 
\end{pf} 
Now, we are ready to establish  the main results of this section, which shows that the value iteration as defined in Eq.~\eqref{eq:valuefunctionateachstage}  indeed characterizes the probability of satisfying the reach-avoid specification in the worst case. 
\begin{mythm}\label{thm:markovoptimal}
Let $\mathbf{v}_0$ be the value function for stage $t=0$ computed by Eq.~\eqref{eq:valuefunctionateachstage}. 
Then for any $x\in X$, it holds that 
\begin{equation}\label{eq:valuestage0isreachavoid}
    \mathbf{v}_0(x) = \mathsf{RA}_x^\star(G,S).
\end{equation} 
Moreover, let $\pi^\star=(\pi_0^\star, \pi_1^\star,\dots,\pi_{T-1}^\star) \in \Pi$ 
be the control policy such that Eq.~\eqref{eq:max-achieve} holds for each $t=0,1,\dots,T-1$. 
Then we have  
\begin{equation}\label{eq:optimalpolicy}
\mathbf{v}_0(x)= \mathsf{RA}_x^{\pi^\star}(G,S), 
\end{equation}
i.e., 
$\pi^\star$ computed is indeed the optimal control policy.  
\end{mythm}
\begin{pf}
For control policy
$\pi=(\pi_0,\pi_1,\dots, \pi_{T-1}) \in \Pi$, 
adversarial strategy 
$\gamma=(\gamma_0,\gamma_1,\dots, \gamma_{T-1}) \in \Gamma$,  
and time instant $t=0,1\dots, T-1$, 
we define a cost-to-go function $V_t^{\pi,\gamma}: X\to [0,1]$ by:
\begin{equation}\label{eq:fixed policy induction}
    \begin{aligned}
       V_t^{\pi,\gamma}&(x_t):= \\
       &\mathbb{E}^{\pi_{\downarrow t},\gamma_{\downarrow t}}_{x_t}\left[ \mathbf{1}_{G}(x_t) + \sum_{j=t+1}^T \left( \prod_{i=t}^{j-1} \mathbf{1}_{S\setminus G}(x_i)\right) \mathbf{1}_G(x_j)\right], 
    \end{aligned}
\end{equation}
where $\pi_{\downarrow t}=(\pi_t,\pi_{t+1},\dots \pi_{T-1})$ and the same for $\gamma_{\downarrow t}$. 
We define $V_T^{\pi,\gamma}(x_T):=\mathbf{1}_G(x_T)$. Clearly, we have $\textsf{RA}_{x_0}^{\pi,\gamma}(G,S)= V_0^{\pi,\gamma}(x_0)$. 
For two  measurable $g: X \to U$ and $h: X \times U \to \mathbb{D}$  functions, 
we define operator $\mathbb{T}_{g,h}$ on the value functions by: $\forall x \in X$,
\begin{equation}\label{eq:fixed policy operator}
        \mathbb{T}_{g,h} (\mathbf{v})(x):=\mathbf{1}_{G}(x) + \mathbf{1}_{S\setminus G}(x) \mathbf{H}(x,g(x),h(x,g(x)),\mathbf{v}).
\end{equation}
Then according to Lemma 2 in \cite{ding2013stochastic}, it holds that 
    \begin{equation} \label{eq:onestep-equation}
        V_t^{\pi,\gamma}(x)=\mathbb{T}_{\pi_t,\gamma_t}(V_{t+1}^{\pi,\gamma})(x), \forall x \in X, t=0,1,\dots,T-1.
    \end{equation}
Now, we consider the distributionally-robust setting. We claim that 
\begin{equation} \label{eq:claim}
\forall t=0,1,\dots,T, \exists \pi^{t} \in \Pi,\forall \gamma \in \Gamma: 
\mathbf{v}_t\leq V_t^{\pi^{t},\gamma}.
\end{equation}
We prove \eqref{eq:claim} by induction on $t$.
When $t=T$, since $\mathbf{v}_T= V_T^{\pi,\gamma}=\mathbf{1}_G$ for any $\pi \in \Pi$, $\gamma \in \Gamma$, the claim is true. Now assume that the claim holds for $t=k+1$. Let $\pi^{k+1} \in \Pi$ be a control policy satisfying the claim when $t=k+1$. By Proposition~\ref{prop:optimalcontrolandsuboptimaladv}-(1), we can find measurable $\pi_k^\star : X \to U$ such that $\pi^\star_k(x) \in \arg \max_{u \in U} \inf_{\mu \in \mathbb{D}} \mathbf{H}(x,u,\mu,\mathbf{v}_{k+1})$. Consider $\pi^{k} \in \Pi$ s.t. $\pi^{k}_{\downarrow k}=(\pi_k^\star,\pi^{k+1}_{\downarrow k+1})$. Then for any $x \in X$ and $\gamma = (\gamma_0,\gamma_1,\dots, \gamma_{T-1}) \in \Gamma$,
\begin{align*}
    &V_k^{\pi^{k},\gamma}(x) =\mathbb{T}_{\pi^\star_k,\gamma_k}(V_{k+1}^{\pi^{k+1},\gamma})(x)  \geq \mathbb{T}_{\pi^\star_k,\gamma_k}(\mathbf{v}_{k+1})(x)\\
      &  =  \mathbf{1}_{G}(x) + \mathbf{1}_{S\setminus G}(x) \mathbf{H}(x,\pi^\star_k(x),\gamma_k(x,\pi^\star_k(x)),\mathbf{v}_{k+1})\\
       & \geq  \mathbf{1}_{G}(x) + \mathbf{1}_{S\setminus G}(x) \inf_{\mu \in \mathbb{D}} \mathbf{H}(x,\pi^\star_k(x),\mu,\mathbf{v}_{k+1})\\
      &  =  \mathbb{T}(\mathbf{v}_{k+1})(x)=\mathbf{v}_k(x).
\end{align*}
The first equality comes from $\pi^{k}_{\downarrow k+1}=\pi^{k+1}_{\downarrow k+1}$, \eqref{eq:fixed policy induction} and \eqref{eq:onestep-equation}. The first inequality holds from monotonicity of the operator $\mathbb{T}_{g,h}$ in \eqref{eq:fixed policy operator} and hypothesis induction.
Thus the claim holds by induction. We obtain $\pi^{0} \in \Pi$ such that
\[
\mathbf{v}_0(x)\leq V_0^{\pi^{0},\gamma}(x)=\textsf{RA}_{x}^{\pi^{0},\gamma}(G,S), \forall x \in X, \gamma \in \Gamma.
\]
By applying infimum over $\Gamma$ in equation above, we have
\begin{equation} \label{eq:middlemiddleleq}
    \mathbf{v}_0(x)\leq \inf_{\gamma \in \Gamma}\textsf{RA}_{x}^{\pi^{0},\gamma}(G,S) \leq \textsf{RA}^\star_x(G,S), \forall x \in X.
\end{equation}
We now prove following claim by induction: 
\begin{equation}
\begin{aligned}
    \forall t = 0,1,\dots, T, &\forall \epsilon >0, \\
    &\exists \gamma^{t,\epsilon} \in \Gamma, \forall \pi \in \Pi : \mathbf{v}_t+\epsilon \geq V_t^{\pi,\gamma^{t,\epsilon}}.
\end{aligned} \nonumber
\end{equation}
When $t=T$, the claim holds. Now assume that the claim holds for $t=k+1$. Let $\epsilon >0$ be a given value and $\gamma^{k+1,\epsilon/2} \in \Gamma$ be an adversary strategy satisfying the claim when $t=k+1$. By Proposition~\ref{prop:optimalcontrolandsuboptimaladv}-(2), we can find measurable $\gamma_k^\star : X \times U \to \mathbb{D}$ such that $\forall (x,u) \in X \times U$, $ \mathbf{H}(x,u,\gamma^\star_t(x,u),\mathbf{v}_{t+1}) \leq \inf_{\mu \in \mathbb{D}} \mathbf{H}(x,u,\mu,\mathbf{v}_{t+1})+\epsilon/2$. Consider $\gamma^k \in \Gamma$ s.t. $\gamma^k_{ \downarrow k}=(\gamma_k^\star,\gamma^{k+1,\epsilon/2}_{\downarrow k+1})$. For any $x \in X$ and $\pi=(\pi_0,\pi_1,\dots,\pi_{T-1}) \in \Pi$,
 \begin{equation}
    \begin{aligned}
        &V_k^{\pi,\gamma^{k}}(x)=\mathbb{T}_{\pi_k,\gamma_k^\star}(V_{k+1}^{\pi,\gamma^{k+1,\epsilon/2}})(x) \\ 
        \leq &\mathbb{T}_{\pi_k,\gamma_k^\star}(\mathbf{v}_{k+1}+\epsilon/2)(x)=  \mathbb{T}_{\pi_k,\gamma_k^\star}(\mathbf{v}_{k+1})(x)+\epsilon/2 \\
        = & \mathbf{1}_{G}(x) + \mathbf{1}_{S\setminus G}(x) \mathbf{H}(x,\pi_k(x),\gamma_k^\star(x,\pi_k(x)),\mathbf{v}_{k+1})+\epsilon/2\\
       \leq &  \mathbf{1}_{G}(x) + \mathbf{1}_{S\setminus G}(x) \inf_{\mu \in \mathbb{D}} \mathbf{H}(x,\pi_k(x),\mu,\mathbf{v}_{k+1})+\epsilon/2+\epsilon/2\\
      \leq & \mathbb{T}(\mathbf{v}_{k+1})(x)+\epsilon=\mathbf{v}_k(x)+\epsilon.
    \end{aligned} \nonumber
\end{equation}
The first equality comes from $\gamma^k_{\downarrow k+1}=\gamma^{k+1,\epsilon/2}_{\downarrow k+1}$, \eqref{eq:fixed policy induction} and \eqref{eq:onestep-equation}. The first inequality holds from monotonicity of the operator $\mathbb{T}_{g,h}$ in \eqref{eq:fixed policy operator} and inductive hypothesis. The second inequality is true from definition of $\gamma^\star_k$. Thus the claim holds by induction. For any $\epsilon > 0$, we obtain $\gamma^{0,\epsilon} \in \Gamma$ such that for any $x\in X$ and $\pi \in \Pi$, 
\[
\mathbf{v}_0(x)+\epsilon \geq V_0^{\pi,\gamma^{0,\epsilon}}(x)=\textsf{RA}_{x}^{\pi,\gamma^{0,\epsilon}}(G,S).
\]
Thus for any $\epsilon >0$, by applying supremum over $\Pi$ in equation above, we have
\[
\mathbf{v}_0(x)+\epsilon \geq \sup_{\pi \in \Pi}\textsf{RA}_{x}^{\pi,\gamma^{0,\epsilon}}(G,S) \geq \textsf{RA}^\star_x(G,S), \forall x \in X.
\]
Since $\epsilon>0$ is arbitrary and \eqref{eq:middlemiddleleq}, for $\pi^0$ in \eqref{eq:middlemiddleleq},
\[
\mathbf{v}_0(x) \leq \textsf{RA}_x^{\pi^0}(G,S) \leq \textsf{RA}^\star_x(G,S) \leq \mathbf{v}_0(x), \forall x \in X.
\]
Thus \eqref{eq:valuestage0isreachavoid} and \eqref{eq:optimalpolicy} hold. This completes the proof.
$\hfill\square$
\end{pf}

Theorem~\ref{thm:markovoptimal} indicates that if $\mathbf{v}_0(x)\geq \alpha$ for all $x \in X_0$, then
$\pi^\star \in \Pi$ computed according to the dynamic program  is indeed a solution of Problem~\ref{problem:threshold}-($\textsf{RA}$). 
However, this result is mainly theoretical as  the optimization problem in Eq.~\eqref{eq:optimalpolicy} is infinite-dimensional. 
To effectively address this computational challenge, in the next subsection,  we will further convert it to  a finite-dimensional optimization problem.

\begin{remark}\label{remark:comparewithreachavoidgame}
In \cite{ding2013stochastic}, the authors also consider the reach-avoid control synthesis problem within a stochastic game framework. Although \cite{ding2013stochastic} addresses a more general stochastic hybrid system than our continuous system, the dynamic game framework in \cite{ding2013stochastic} is inapplicable to our work. Specifically, \cite{ding2013stochastic} assumes that for every Borel subset of the state space, the stochastic kernel is continuous with respect to the adversary variable. This assumption does not hold in our setting, as it conflicts with the convergence of probability measures in the weak topology \cite[Sec. 7.4.2]{bertsekas1996stochastic}. Moreover, due to the differences in game frameworks, while the optimal adversary policy always exists in \cite{ding2013stochastic}, our work (as stated in Proposition 2-(2)) guarantees only the existence of a sub-optimal policy for the adversary.
\end{remark}

\subsection{Computation  Considerations for Optimal Policies}\label{subsection:dualmethod}
In general, solving the Bellman equation~\eqref{eq:operator-1} to evaluate $\mathbf{v}_t(x)$ with the Wasserstein ambiguity set \eqref{eq:ambiguitydef} is challenging, as it involves infinite-dimensional min-max optimization problems. 
To address this difficulty, we first reformulate the optimization problem in the dynamic programming for each stage of value function using duality result in~\cite{gao2023distributionally}.
 
\begin{mythm}\label{thm:dualsolution}
 For any $x \in X$ and $t=0,1,\dots,T-1$, it holds that 
    \begin{equation}\label{eq:dualoptimization}
    \begin{aligned}
        \mathbf{v}_t(x)& = \mathbf{1}_{G}(x) + \mathbf{1}_{S\setminus G}(x) \times  \max_{\lambda \geq 0 ,u \in U} \bigg\{ -\lambda \theta^p+ \\
        & \int_{W} \inf_{w \in W}\left[ \mathbf{v}_{t+1}(f(x,u,w))+\lambda d^p(w,\hat{w}) \right] \nu_N(\mathrm{d}\hat{w}) \bigg\} 
    \end{aligned}
    \end{equation}
 with $\mathbf{v}_T=\mathbf{1}_G$.
 Furthermore, let $u_t^\star(x)$ be the optimal solution to the optimization problem \eqref{eq:dualoptimization} for $(x,t)$. 
 Then for policy $u^\star=(u_0^\star,u_1^\star,\dots,u_{T-1}^\star) \in \Pi$, we have $\mathsf{RA}_{x}^{u^\star}(G,S)= \mathsf{RA}^\star_{x}(G,S)$ for all $x \in X$.
\end{mythm}
\begin{pf} 
Let $V:  W \to \mathbb{R}$ be a value function. Consider the following primal and dual problems:
\begin{equation}\label{eq:wasprimaldual}
 \begin{aligned}
     v_P&=\inf_{\mu \in \mathcal{P}(W)} \left\{ \int_{W} V(w) \mu(\mathrm{d}w) \mid \mathcal{W}_p(\mu,\nu_{N}) \leq \theta \right\},  \\
     v_D 
     &= \sup_{\lambda \geq 0}\left[ -\lambda \theta^p+\int_{W}\inf_{w \in W}\left[ V(w)+\lambda d^p(w,\hat{w}) \right]\nu_N(\mathrm{d}\hat{w}) \right].
 \end{aligned}   
\end{equation}
According to  \cite{gao2023distributionally}, for any bounded $W$ and  measurable and bounded function $V: W\to \mathbb{R}$,  
the dual problem in \eqref{eq:wasprimaldual} always admits a maximum $\lambda^\star$ and $v_P=v_D<\infty$.
Therefore, for each $x \in X$ and $t=0,1\dots,T-1$, we have
\begin{equation}\label{eq:provemiddle}
    \begin{aligned}
       \sup_{u \in U}&\inf_{\mu 
    \in \mathbb{D}} \mathbf{H}(x,u,\mu,\mathbf{v}_{t+1})=\sup_{\lambda \geq 0 ,u \in U} \bigg\{-\lambda \theta^p+  \\
        & \int_{W} \inf_{w \in W}\left[ \mathbf{v}_{t+1}(f(x,u,w))+\lambda d^p(w,\hat{w}) \right] \nu_N(\mathrm{d}\hat{w}) \bigg\}.
    \end{aligned}
\end{equation}
From Proposition~\ref{prop:optimalcontrolandsuboptimaladv}-(1), the supremum in \eqref{eq:provemiddle} can be replaced by maximum. 
Thus \eqref{eq:dualoptimization} holds. Then by Theorem~\ref{thm:markovoptimal}, $\textsf{RA}_{x}^{u^\star}(G,S)= \textsf{RA}^\star_{x}(G,S)$ for all $x \in X$. $\hfill\square$
\end{pf}

In general, the above dual optimization problem is still infinite-dimensional. 
However, by further leveraging the assumption that  the nominal distribution $\nu_N$ has finite support in the form of  Eq.\eqref{eq:finitesupportdistribution},  the optimization problem in Eq.\eqref{eq:dualoptimization} is equivalent to the following program: 
\begin{align} 
     &\max_{u\in U, \lambda\geq 0, l_i\in \mathbb{R}} \quad  -\lambda \theta^p+\sum_{i=1}^{M} p_i l_i\label{eq:finitesupportdual}\\
      & \text{s.t.}  \quad   l_i \leq \mathbf{v}_{t+1}(f(x,u,w))+\lambda d^p(w,\hat{w}_i),\forall w \in W.  \nonumber
\end{align}
Intuitively, when the nominal distribution has finite support, Eq.\eqref{eq:finitesupportdual} replaces the integral in Eq.\eqref{eq:dualoptimization} with a probabilistic weighted sum, where each variable  $l_i$ represents the infimum value when $\hat{w}=\hat{w}_i$.
Note that the variables $\lambda$ and $u$ in Eq.~\eqref{eq:finitesupportdual} may have multiple optimal solutions. Moreover, every optimal solution $u$ corresponds to the optimal control policy at time instant $t$ and state $x$, i.e., the policy $\pi_t(x)$.
The optimization problem in Eq.~\eqref{eq:finitesupportdual} is a semi-infinite optimization program for which various convergent methods exist, such as discretization methods, primal-dual methods, and constraint sampling methods \cite{garatti2024non,doi:10.1137/1035089,lopez2007semi}.
In our later experiments, we employ the discretization algorithm from \cite{reemtsen1991discretization}, which adaptively generates grids over $W$. This approach ensures convergence to a local optimal solution of the semi-infinite program. 
Note that the complexity of computing $\mathbf{v}_t$ is exponential in the dimension of $X$, as solving via dynamic programming requires discretizing $X$. Additionally, in the more complex distributionally robust setting, when solving Eq.~\eqref{eq:finitesupportdual} for a specific $x \in X$, the number of constraints typically grows exponentially with the dimension of the disturbance space $W$. This complexity is difficult to avoid, as the optimization involves general non-convex and potentially discontinuous functions $\mathbf{v}_t$. To alleviate this exponential complexity, in the next section, we will demonstrate how to compute an approximate solution to the safety control synthesis problem.
\begin{remark}
The  control policy $u^\star \in \Pi$ computed  in Theorem~\ref{thm:dualsolution} actually achieves the optimal reach-avoid specification probability.
In Problem~\ref{problem:threshold}, we only require that the satisfaction probability exceeds the  threshold $\alpha$. 
In this case, for each state   $x \in X$ and time instant $t=0,1\dots,T-1$, it suffices to pick an arbitrary control input from the following set
    \begin{equation}\label{eq:feasiblecontrol}
         \mathcal{U}_t(x)=\{ u \in U \mid \inf_{\mu \in \mathbb{D}}\mathbf{H}(x,u,\mu, \mathbf{v}_{t+1}) \geq \alpha  \}
    \end{equation}
as long as it holds that $\mathbf{v}_t(x)\geq \alpha$.  
\end{remark}
\begin{remark}
Here we discuss how $\theta$ the satisfaction probability, which is equivalent to the objective value of Eq.~\eqref{eq:finitesupportdual}.  When $\theta = 0$, the Wasserstein ambiguity set only contains the distribution $\sum_{i=1}^M p_i \delta_{\hat{w}_{i}}$. In this case, the optimal value of Eq.~\eqref{eq:finitesupportdual} is$\sum_{i=1}^M p_i \mathbf{v}_{t+1}(f(x,u,w_i))$, since $\lambda$ should be set to infinity. When $\theta=\infty$, the Wasserstein ambiguity set contains all probability measures on $W$, and the optimal value of Eq.~\eqref{eq:finitesupportdual} is $\inf_{w \in W}\{\mathbf{v}_{t+1}(f(x,u,w))\}$, since $\lambda$ should be set to zero in this case.
Therefore, as $\theta$  increases, the Wasserstein ambiguity set includes more distributions, and the computed reach-avoid probability becomes smaller.
We refer readers to~\cite{gao2023distributionally} for more details on how to choose $\theta$ such that the true distribution is contained within the ambiguity set $\mathbb{D}$ with a given confidence, while ensuring that the distributionally robust optimization is not overly conservative.    
\end{remark}
\subsection{Case of Safety Specifications}\label{subsection:safetysolution}
In this subsection, we  extend the above results to  solve Problem~\ref{problem:threshold}-(\textsf{SA}).
Let $\mathbf{v}$ be a measurable function on $X$.
We define another operator $\hat{\mathbb{T}}$ by: 
for any $x\in X$, we have
\begin{equation}\label{eq:safetyoperator}
\hat{\mathbb{T}}(\mathbf{v})(x):=\mathbf{1}_{S}(x) \sup_{u \in U} \inf_{\mu 
    \in \mathbb{D}} \mathbf{H}(x,u,\mu,\mathbf{v}), 
\end{equation}
where $\mathbf{H}(x,u,\mu,\mathbf{v})$ is the integration in \eqref{eq:operator-2}. 
Similar to \eqref{eq:valuefunctionateachstage}, we define a sequence of value functions  $\{\hat{\mathbf{v}}_t\}_{t=0,1,\dots,T}$ by:
\begin{align} \label{eq:safetyvaluefunctionateachstage}
		\left\{
		\begin{array}{ll }
  \hat{\mathbf{v}}_T=\mathbf{1}_S\\
		\hat{\mathbf{v}}_t=\hat{\mathbb{T}}(\hat{\mathbf{v}}_{t+1}), \forall t=0,1,\dots, T-1
		\end{array}. 
		\right.   
\end{align}
Based on this new operator, we can establish the following results for the case of safety specifications, analogous to Theorems~\ref{thm:markovoptimal} and \ref{thm:dualsolution}, which address the case of reach-avoid specifications. 
The proofs are omitted here because the operator $\mathbb{T}$ is already more complex than $\hat{\mathbb{T}}$. All proofs for  reach-avoid specification  can be extended to   safety specification  without loss of generality.

\begin{mycol}\label{col:markovoptimalforsafety}
Let $\hat{\mathbf{v}}_0$ be the value function for stage $t=0$ computed by Eq.~\eqref{eq:safetyvaluefunctionateachstage}. 
Then for any $x\in X$, it holds that
$\hat{\mathbf{v}}_0(x)=\mathsf{SA}^\star_{x}(S)$.
Furthermore, let $\pi^\star=(\pi_0^\star, \pi_1^\star,\dots,\pi_{T-1}^\star) \in \Pi$ 
be the control policy such that
$\pi^\star_t(x) \in \arg \max_{u \in U} \inf_{\mu 
    \in \mathbb{D}} \mathbf{H}(x,u,\mu,\hat{\mathbf{v}}_{t+1})$.
Then we have 
$\mathsf{SA}_{x}^{\pi^\star}(S)= \mathsf{SA}^\star_{x}(S)$.
\end{mycol}

Similarly, the  optimal policy for Problem~\ref{problem:threshold}-(\textsf{SA}) can also be computed by the dual method. 
 
\begin{mycol}\label{col:dualmethodsafety}
For any $x \in X$ and $t=0,1,\dots,T-1$, it holds that 
\begin{equation}\label{eq:safetydualoptimization}
    \begin{aligned}
        \hat{\mathbf{v}}_t(x)& =  \mathbf{1}_{S}(x) \times  \max_{\lambda \geq 0 ,u \in U} \bigg\{ -\lambda \theta^p+ \\
        & \int_{W} \inf_{w \in W}\left[ \hat{\mathbf{v}}_{t+1}(f(x,u,w))+\lambda d^p(w,\hat{w}) \right] \nu_N(\mathrm{d}\hat{w}) \bigg\} 
    \end{aligned}
    \end{equation}
 with $\mathbf{v}_T=\mathbf{1}_S$. 
 Furthermore, let $u_t^\star(x)$ be the optimal solution to the optimization problem \eqref{eq:safetydualoptimization} for $(x,t)$. 
Then for policy $u^\star=(u_0^\star,u_1^\star,\dots,u_{T-1}^\star) \in \Pi$, we have $\mathsf{SA}_{x}^{u^\star}(S)= \mathsf{SA}^\star_{x}(S)$ for all $x \in X$.  
\end{mycol}

\begin{remark}\label{remark:comparewithdrsafety}
In \cite{yang2018dynamic}, the author also addresses the distributionally robust control synthesis problem for safety specifications. However,  the results in \cite{yang2018dynamic} rely on the additional assumption that, for any bounded continuous function
$g: \mathbb{R}^n\to \mathbb{R}$, function $\int_{W}g(f(x,u,w))\emph{d}\mu(w)$ is continuous in  $(x,u,\mu) \in X \times U \times \mathbb{D}$.  This assumption holds when $X$,   $U$ and  $W$ are all compact.  
However, for example, when $X=\mathbb{R}^n$, this assumption does not hold. Our results establish the upper semi-continuity of  $\hat{\mathbf{v}}_t$ in Eq.~\eqref{eq:safetyoperator}, thereby relaxing the condition stated in Theorem 1 of \cite{yang2018dynamic}. 
Furthermore, our results are developed based on the more complex reach-avoid specification that is not studied in \cite{yang2018dynamic}.
\end{remark}

\section{Safe Control Synthesis using Distributionally Robust Control Barrier Certificates}\label{sec:approximate}
In practice, explicitly solving the dual optimization problem discussed in the previous section remains computationally challenging for high-dimensional systems. To address this, in this section, we focus exclusively on safety specifications and propose an approach to ensure a lower bound on the satisfaction probability using a novel concept of \emph{distributionally robust control barrier certificates} (DR-CBC). For polynomial systems, we further demonstrate that the control synthesis problem can be solved more efficiently by leveraging the sum-of-squares (SOS) programming technique.

\subsection{Distributionally Robust   Control Barrier Certificates}\label{subsection:derivateappro}  
First, we formally define distributionally robust control barrier certificates, which provide a characterization of the lower bound on the safety probability and the associated control policy.

\begin{mydef}\label{def:DRSCBC}
Given  system $\mathbb{S}=(X,X_0,U,W,f,T)$ and safety set $S \subseteq X$, 
function $\Bar{\mathbf{v}}: X \to \mathbb{R}$ is said to be a  
\emph{distributionally robust   control barrier certificate} (DR-CBC) if the following two conditions hold:\vspace{-3pt}
    \begin{enumerate}
        \item 
        $\Bar{\mathbf{v}} \leq \mathbf{1}_S$; \medskip
        \item 
        There exists $\beta \leq 0$ and $\eta > 0$ such that
       \begin{equation}\label{eq:decrease}
       \beta\leq \inf_{x \in S} \left[ \sup_{u \in U} \inf_{\mu 
    \in \mathbb{D}} \mathbf{H}(x,u,\mu,\Bar{\mathbf{v}}) - \frac{\Bar{\mathbf{v}}(x)}{\eta}\right].
        \end{equation}
\end{enumerate} 
Furthermore, a function $\mathbf{u}:X\to U$ is said to be  the \emph{associated control function} w.r.t. DR-CBC $\Bar{\mathbf{v}}$ if it achieves the inequality in Eq.~\eqref{eq:decrease}, i.e., 
\begin{equation}\label{eq:corresponging control for CBC}
       \beta\leq \inf_{x \in S} \left[ \inf_{\mu 
    \in \mathbb{D}} \mathbf{H}(x,\mathbf{u}(x),\mu,\Bar{\mathbf{v}}) - \frac{\Bar{\mathbf{v}}(x)}{\eta} \right].
\end{equation} 
We denote $\pi_{\mathbf{u}}=(\mathbf{u},\mathbf{u},\dots,\mathbf{u})\in \Pi$ as the \emph{associated control policy}, which applies  $\mathbf{u}$ at each time step.
\end{mydef}

The intuitions of the DR-CBC $\Bar{\mathbf{v}}$ and its associated control function $\mathbf{u}$ can be interpreted as follows. 
Let us define operator by
\begin{equation}\label{eq:operatorwithcorres}
\hat{\mathbb{T}}_{\mathbf{u}}(\mathbf{v})(x):= \mathbf{1}_{S}(x)  \inf_{\mu 
    \in \mathbb{D}} \mathbf{H}(x,\mathbf{u}(x),\mu,\mathbf{v}), \forall x \in X,
\end{equation}
which modifies the operator $\hat{\mathbb{T}}$ in Eq.~\eqref{eq:safetyoperator} by fixing control inputs according to the given $\mathbf{u}$, and define value functions 
 $\{\hat{\mathbf{v}}_t^{\mathbf{u}}\}_{t=0,1,\dots,T}$ and $\{\Bar{\mathbf{v}}_t^{\mathbf{u}}\}_{t=0,1,\dots,T}$ by 
\begin{align}   
		\left\{
		\begin{array}{ll }
\hat{\mathbf{v}}^{\mathbf{u}}_T&=\mathbf{1}_S,\\ 
\hat{\mathbf{v}}^{\mathbf{u}}_t&=\hat{\mathbb{T}}_{\mathbf{u}}(\hat{\mathbf{v}}^{\mathbf{u}}_{t+1})
		\end{array}
		\right. 
\text{ and }\quad   
		\left\{
		\begin{array}{ll }
\Bar{\mathbf{v}}^{\mathbf{u}}_T&=\Bar{\mathbf{v}},\\
\Bar{\mathbf{v}}^{\mathbf{u}}_t&=\hat{\mathbb{T}}_{\mathbf{u}}(\Bar{\mathbf{v}}^{\mathbf{u}}_{t+1})
		\end{array}
		\right..  
\end{align}
As shown in Corollary~\ref{col:markovoptimalforsafety}, \( \hat{\mathbf{v}}_0^{\mathbf{u}} \)    encapsulates information about the safety probability when the system adopts the associated control policy $\pi_{\mathbf{u}}$. Moreover, \( \Bar{\mathbf{v}} \) essentially provides a lower bound for 
\(\hat{\mathbf{v}}_0^{\mathbf{u}} \). 
To see this, condition (1) in Definition~\ref{def:DRSCBC} ensures that $\Bar{\mathbf{v}}_T^{\mathbf{u}} \leq \hat{\mathbf{v}}_T^{\mathbf{u}}$. 
Furthermore, by the monotonicity of operator $\hat{\mathbb{T}}_{\mathbf{u}}$, we have $\Bar{\mathbf{v}}_t^{\mathbf{u}} \leq \hat{\mathbf{v}}_t^{\mathbf{u}}$ for each stage $t$.
Moreover, value $\beta$ in condition (2) provides a lower bound on decrease rate of $\Bar{\mathbf{v}}$ after applying operator $\hat{\mathbb{T}}_{\mathbf{u}}$.  
Thus, we can use \( \beta \) and \( \eta \) to recursively establish a lower bound for each \( \Bar{\mathbf{v}}_t^{\mathbf{u}} \) in terms of \( \Bar{\mathbf{v}} \). 
The above intuition is formalized by the following theorem.   
\begin{mythm}\label{thm:approximatereachavoidpro}
Given  system $\mathbb{S}=(X,X_0,U,W,f,T)$ and safety set $S \subseteq X$,  suppose that $\Bar{\mathbf{v}}:X\to \mathbb{R}$ is a DR-CBC with parameters $\beta \leq 0$ and $\eta > 0$. Then for its associated control policy $\pi_{\mathbf{u}}\in \Pi$, we have
\begin{equation}\label{eq:approximateresult}
\mathsf{SA}_{x_0}^{\pi_{\mathbf{u}}}(S) \geq \eta^{-T}\Bar{\mathbf{v}}(x_0) + (\sum_{i=0}^{T-1} \eta^{-i})\beta.
\end{equation}
\end{mythm}
\begin{pf}
We claim, for $t=0,1,\dots,T$, it holds that \vspace{-3pt}
    \begin{enumerate}
        \item $\Bar{\mathbf{v}}^{\mathbf{u}}_t \leq \hat{\mathbf{v}}^{\mathbf{u}}_t$;\medskip
        \item $ \eta^{t-T}\Bar{\mathbf{v}}  + (\sum_{i=0}^{T-t-1} \eta^{-i}) \beta \leq \Bar{\mathbf{v}}^{\mathbf{u}}_t$.
        \vspace{-3pt}
    \end{enumerate}
 We prove (1) by induction. For $n=T$, $\Bar{\mathbf{v}}^{\mathbf{u}}_T =  \Bar{\mathbf{v}}\leq \mathbf{1}_S = \hat{\mathbf{v}}^{\mathbf{u}}_T$. Thus (1) holds for $n=T$. Now suppose that (1) holds for $n=t+1$ and consider case when $n=t$. For $x \in X \setminus S$, we have $\Bar{\mathbf{v}}^{\mathbf{u}}_t(x) = \hat{\mathbf{v}}^{\mathbf{u}}_t(x)= 0$. For $x \in S$, we have $\Bar{\mathbf{v}}^{\mathbf{u}}_t(x)= \inf_{\mu 
    \in \mathbb{D}} \mathbf{H}(x,\mathbf{u}(x),\mu,\Bar{\mathbf{v}}^{\mathbf{u}}_{t+1})  \leq \inf_{\mu 
    \in \mathbb{D}} \mathbf{H}(x,\mathbf{u}(x),\mu,\hat{\mathbf{v}}^{\mathbf{u}}_{t+1})=\hat{\mathbf{v}}^{\mathbf{u}}_t(x)$.
The inequality is true due to inductive hypothesis $\Bar{\mathbf{v}}^{\mathbf{u}}_{t+1}\leq \hat{\mathbf{v}}^{\mathbf{u}}_{t+1}$ and monotonicity of the operator $\hat{\mathbb{T}}_{\mathbf{u}}$. Therefore (1) holds for $n=t$. This completes the Proof of (1).

We still prove (2) by induction. For $n=T$, (2) is true from definition of $\Bar{\mathbf{v}}^{\mathbf{u}}_T$ and $\sum_{i=0}^{-1} \eta^{-i}=0$. Now suppose that (2) holds for $n=t+1$ and consider case when $n=t$. For $x \in X \setminus S$, we have
\[
\eta^{t-T}\Bar{\mathbf{v}}(x)  + (\sum_{i=0}^{T-t-1} \eta^{-i}) \beta \leq  \eta^{t-T} \Bar{\mathbf{v}}(x) \leq 0= \Bar{\mathbf{v}}_t^{\mathbf{u}}(x).
\]
The first inequality holds since $(\sum_{i=0}^{T-t-1} \eta^{-i}) \beta \leq 0$. The second inequality comes from $\Bar{\mathbf{v}}(x)\leq \mathbf{1}_S(x)=0$ and $\eta^{t-T}>0$. The equality comes from definition of $\hat{\mathbb{T}}_{\mathbf{u}}$ in \eqref{eq:operatorwithcorres}. For $x \in S$,
\begin{align*}
    &\Bar{\mathbf{v}}_t^{\mathbf{u}}(x) =\hat{\mathbb{T}}_{\mathbf{u}}(\Bar{\mathbf{v}}^{\mathbf{u}}_{t+1})(x) \geq \hat{\mathbb{T}}_{\mathbf{u}}(\eta^{t+1-T}\Bar{\mathbf{v}})(x)  + (\sum_{i=0}^{T-t-2} \eta^{-i}) \beta\\ 
   &  = \eta^{t+1-T}\hat{\mathbb{T}}_{\mathbf{u}}(\Bar{\mathbf{v}})(x)+(\sum_{i=0}^{T-t-2} \eta^{-i}) \beta \geq \eta^{t+1-T}(\frac{\Bar{\mathbf{v}}(x)}{\eta} +\beta) \\
    &  +(\sum_{i=0}^{T-t-2} \eta^{-i}) \beta=\eta^{t-T}\Bar{\mathbf{v}}(x) +(\sum_{i=0}^{T-t-1} \eta^{-i}) \beta. 
\end{align*}
The first inequality comes from inductive hypothesis and monotonicity of $\hat{\mathbb{T}}_{\mathbf{u}}$. The second equality holds from definition of $\hat{\mathbb{T}}_{\mathbf{u}}$. The last inequality holds since \eqref{eq:decrease} is true and $\mathbf{u}$ is a associated control function. Thus the claim is true and $\textsf{SA}^{\pi_{\mathbf{u}}}_{x_0}(S) = \hat{\mathbf{v}}^{\mathbf{u}}_0(x_0) \geq \eta^{-T}\Bar{\mathbf{v}}(x_0) + (\sum_{i=0}^{T-1} \eta^{-i})\beta$ where the equality holds from Corollary~\ref{col:markovoptimalforsafety}. $\hfill\square$
\end{pf}
\begin{remark}
It is worth noting that our DR-CBC is defined in a static fashion, with the associated control policy  $\pi_{\mathbf{u}}$ being time-invariant. This idea can, in principle, be extended to time-varying control policies. In this case, the DR-CBC would be defined as a sequence of functions 
$\{\Bar{\mathbf{v}}_t\}_{t=0,1,\dots,T-1}$ such that $\Bar{\mathbf{v}}_0 \leq \cdots \leq \Bar{\mathbf{v}}_{T-1} \leq \mathbf{1}_S$.  
The associated control policy 
$\{\mathbf{u}_t\}_{t=0,1,\dots,T-1}$ would then consist of a sequence of time-varying functions, where for each stage, $\mathbf{u}_t$  satisfies its respective inequality with respect to 
$\Bar{\mathbf{v}}_t $, $\beta_t$, and $\eta_t$.  
In this work, we do not explore this direction further, as it significantly increases the complexity of search for DR-CBCs, which are intended to provide computational efficiency. Moreover, since we are considering safety specifications, employing time-invariant control policies in practice is sufficient and does not introduce significant conservatism.  
\end{remark}
\subsection{Searching DR-CBC via SOS Programs}\label{subsection:certificateapproximate} 
In general, proving the existence of a DR-CBC and finding one remain challenging, as it ultimately reduces to explicitly solving the dynamic program. In this subsection, for a class of polynomial systems, we demonstrate how to search for a DR-CBC and its associated control policy using sum-of-squares (SOS) optimization techniques~\cite{sturm1999using, prajna2002introducing}.

Given \( z \in \mathbb{R}^d \),  let \( \mathbb{R}[z] \) denote the set of all polynomials over \( z \). 
The set of SOS polynomials over $z$ is 
\begin{equation}\label{def:polynomial}
    \Sigma[z] := \left\{ s(z) \in \mathbb{R}[z] \mid s(z)= \sum_{i=1}^m g_i(z)^2, g_i(z) \in \mathbb{R}[z] \right\}. \nonumber
\end{equation}
In order to apply SOS optimization techniques, 
we further make the following assumptions.
\begin{assumption}\label{assumption:polynomial}
The following conditions hold for system $\mathbb{S}=(X,X_0,U,W,f,T)$:\vspace{-3pt}
\begin{enumerate}
    \item 
    $f:X \times U \times W \to X$ is a polynomial;
    \vspace{3pt}
    \item 
    $U \subseteq \mathbb{R}^m$ is a polytope defined by $U=\{ u \in \mathbb{R}^m \mid \mathbf{A}u \geq \mathbf{b} \}$ with $\mathbf{A}=(\mathbf{a})_{ij} \in \mathbb{R}^{r \times m}$ and $\mathbf{b} \in \mathbb{R}^r$; 
    \vspace{3pt}
    \item  
    Each set $\star \in \{X_0,X,S,X \setminus S,S\times U \times W\}$ is defined as the super-level set of polynomial $\mathbf{s}_{\star}$, 
    i.e., $\star=\{ x \in \mathbb{R}^n \mid \mathbf{s}_{\star}(x) \geq 0 \}$. For simplicity, we denote $\mathbf{s}_{S\times U \times W}(x)$
    as $\mathbf{s}_{\zeta}(x)$; \vspace{3pt} 
   \item  
   For the Wasserstein metric in \eqref{eq:wasserdef}, metric $d(w,w')$ and order $p$ satisfy that $d^p(w,w') \in \mathbb{R}[w]$. 
\end{enumerate}
\end{assumption}
The conditions (1)-(3) above are common assumptions for searching barrier certificates by SOS programs. 
Condition (4) is an additional requirement for DR-CBC. 
Specifically, we will express the requirement \eqref{eq:decrease} by the dual representation of the Wasserstein distributionally robust optimization in Corollary~\ref{col:dualmethodsafety}, which contains item $d^p(w,w')$. 
Therefore,  (4) is required in order to apply the SOS program techniques. 
This assumption can be satisfied by choosing, e.g.,  $d(w,w')=\Vert w-w'\Vert_2$ and $p=2$.

 Under Assumption~\ref{assumption:polynomial}, we propose the SOS program~\eqref{eq:firstSOS}-\eqref{eq:seventhSOS} to search the barrier certificate. Specifically, $\Tilde{\mathbf{v}}$ represents the computed DR-CBC, and $\mathbf{u} = [\mathbf{u}_1; \mathbf{u}_2; \dots; \mathbf{u}_m]$ denotes the associated control function.  
 Constraints \eqref{eq:firstSOS} and \eqref{eq:secondSOS} ensure that $\Tilde{\mathbf{v}} \leq \mathbf{1}_S$, while constraint \eqref{eq:thirdSOS} guarantees that $\Tilde{\mathbf{v}}(x) \geq \delta$ for all $x \in X_0$. 
 Constraints \eqref{eq:forthSOS}-\eqref{eq:sixthSOS} leverage the dual representation of the Wasserstein distributionally robust optimization from Corollary~\ref{col:dualmethodsafety} to enforce the satisfaction of Eq.~\eqref{eq:decrease}.  
 In particular, $\lambda(x)$ and $l_i(x)$ in constraint \eqref{eq:forthSOS} approximate the variables $\lambda$ and $l_i$ in Eq.~\eqref{eq:finitesupportdual}, where the non-negativity of  $\lambda(x)$ is enforced by  constraint \eqref{eq:sixthSOS}.
  Then the expression in second line of \eqref{eq:forthSOS} is a lower bound of the objective value of \eqref{eq:finitesupportdual} and \eqref{eq:fifthSOS} represents the constraint of \eqref{eq:finitesupportdual}. 
Finally, the constraint \eqref{eq:seventhSOS} ensures that the stacked function is a well-defined control function. 
 \begin{tcolorboxenv}[t]
\begin{center}
        \tcbset{title=\quad SOS Program for Searching DR-CDC}
    \begin{tcolorbox} 
        \vspace{-15pt}
\begin{align}
 -\Tilde{\mathbf{v}}(x)-\xi_{X \setminus S}(x) \mathbf{s}_{X \setminus S}(x) & \in \Sigma[x] \label{eq:firstSOS} \\
 -\Tilde{\mathbf{v}}(x)-\xi_S(x) \mathbf{s}_{S}(x) +1 & \in \Sigma[x] \label{eq:secondSOS} \\
  \Tilde{\mathbf{v}}(x)-\xi_{X_0}(x) \mathbf{s}_{X_0}(x) -\delta & \in \Sigma[x] \label{eq:thirdSOS} \\
 - \Tilde{\mathbf{v}}(x)/\eta-\xi_{S}(x) \mathbf{s}_{S}(x)-\beta &    \nonumber   \\ 
 -\theta^p \lambda(x)+ \sum_{i=1}^M p_il_i(x)  &\in \Sigma[x] \label{eq:forthSOS} \\
 -l_i(x) -\xi_{\zeta}(x,u,w) \mathbf{s}_{\zeta}(x,u,w)+\lambda&(x) d^p(w,\hat{w}_i) \nonumber \\
 +\Tilde{\mathbf{v}}(f(x,u,w))-\sum_{j=1}^m\left(u_j - \mathbf{u}_j(x)\right) &\in \Sigma[x,u,w] \nonumber \\
          i=1,2,& \dots, M   \label{eq:fifthSOS}\\
\lambda(x) -\xi_{S}(x) \mathbf{s}_{S}(x) & \in \Sigma[x] \label{eq:sixthSOS}\\
\sum_{j=1}^m \mathbf{a}_{kj}\mathbf{u}_j(x)-\xi_{X}(x)\mathbf{s}_{X}(x)-\mathbf{b}_k &\in \Sigma[x] \nonumber \\
k=1,2,&\dots, r\label{eq:seventhSOS}
\end{align} 
 \end{tcolorbox}
\end{center}
\end{tcolorboxenv}    

\begin{mythm}\label{thm:barrier}
Suppose that Assumption \ref{assumption:polynomial} holds 
and let $\Tilde{\mathbf{v}}(x), \lambda(x), l_i(x), \mathbf{u}_j(x) \in \mathbb{R}[x],\xi_{X_0}(x),\xi_{X}(x)$, $\xi_{X\setminus S}(x)$, $ \xi_{S}(x)\in \Sigma[x]$, $\xi_{\zeta}(x,u,w) \in \Sigma[x,u,w]$ be  solutions to SOS program \eqref{eq:firstSOS}-\eqref{eq:seventhSOS}. 
Then for control policy $\pi_{\mathbf{u}}=(\mathbf{u},\mathbf{u},\dots\mathbf{u}) \in \Pi$,
\begin{equation}
\forall x \in X_0: \mathsf{SA}_{x}^{\pi_{\mathbf{u}}}(S)\geq \eta^{-T}\delta+(\sum_{i=0}^{T-1} \eta^{-i})\beta .
\end{equation}
\end{mythm}
\begin{pf}
We first prove that the computed polynomial $\Tilde{\mathbf{v}}(x) \in \mathbb{R}[x]$ is a DR-CBC. We show that the satisfaction of \eqref{eq:firstSOS} implies satisfaction of $\Tilde{\mathbf{v}}(x) \leq 0, \forall x \in X \setminus S$. Specifically, if \eqref{eq:firstSOS} holds, then for $x \in X \setminus S$, we have
 \[
 \begin{aligned}
   -\Tilde{\mathbf{v}}(x) \geq  -\Tilde{\mathbf{v}}(x)-\xi_{X\setminus S}(x) \mathbf{s}_{X \setminus S}(x)\geq 0.
 \end{aligned}
 \]
 The first inequality is true since $\mathbf{s}_{X\setminus S}(x)\xi_{X\setminus S}(x) \geq 0$ holds for all $x \in X \setminus S$. Similarly, we can prove that if \eqref{eq:secondSOS}; \eqref{eq:thirdSOS}; \eqref{eq:seventhSOS} hold, then $\Tilde{\mathbf{v}}(x) \leq 1, \forall x \in S$; $\Tilde{\mathbf{v}}(x) \geq \delta, \forall x \in X_0$; $\mathbf{u}(x)=[\mathbf{u}_1(x);\mathbf{u}_2(x);\dots;\mathbf{u}_m(x)] \in U, \forall x \in X$ hold, respectively. We now prove that if \eqref{eq:forthSOS}-\eqref{eq:sixthSOS} hold, then \eqref{eq:decrease} holds and $\mathbf{u}$ is a associated control function. We prove it by showing that
\begin{equation}\label{eq:SOSmiddle1}
     \inf_{\mu \in \mathbb{D}} \mathbf{H}(x,\mathbf{u}(x),\mu,\Tilde{\mathbf{v}}) - \frac{\Tilde{\mathbf{v}}(x)}{\eta}  \geq \beta, \forall x \in S.
 \end{equation}
 By result of Corollary \ref{col:dualmethodsafety}, \eqref{eq:SOSmiddle1} holds if and only if $\forall x \in S$,
\begin{equation}\label{eq:approxiamtelasttarget}
       \max_{\lambda^\star_x \geq 0} \left\{ -\theta^p \lambda^\star_x+ \sum_{i=1}^M p_i l_i^\star(x, \lambda^\star_x) \right\} -  \frac{\Tilde{\mathbf{v}}(x)}{\eta} \geq \beta, \nonumber
 \end{equation}
 \begin{equation}
     l_i^\star(x,\lambda^\star_x)=\inf_{w \in W} \Tilde{\mathbf{v}}(f(x,\mathbf{u}(x),w))+\lambda^\star_x d^p(w,\hat{w}_i). \nonumber
 \end{equation}
 The satisfaction of \eqref{eq:fifthSOS} implies that, for any $(x,w) \in S \times W$, it holds that
 \begin{equation}
     l_i(x) \leq \Tilde{\mathbf{v}}(f(x,\mathbf{u}(x),w))+\lambda(x)d^p(w,\hat{w}_i). \nonumber
 \end{equation}
By applying infimum over $W$, we have 
 \begin{equation}\label{eq:SOSmiddle2}
 \begin{aligned}
   &\forall x \in S,\quad l_i(x) = \inf_{w \in W} l_i(x) \leq  \\
  & \inf_{w \in W}\Tilde{\mathbf{v}}(f(x,\mathbf{u}(x),w))+\lambda(x) d^p(w,\hat{w}_i) = l_i^\star(x,\lambda(x)).
\end{aligned}
 \end{equation}
The satisfaction of \eqref{eq:forthSOS} implies that
\begin{equation}\label{eq:SOSmiddle3}
     -\theta^p \lambda(x)+ \sum_{i=1}^M p_i l_i(x)  - \frac{\Tilde{\mathbf{v}}(x)}{\eta}\geq \beta, \forall x \in S.
\end{equation}
Then satisfaction of \eqref{eq:forthSOS}-\eqref{eq:sixthSOS} implies that
\begin{align}
    &\sup_{\lambda^\star_x \geq 0} \left\{ -\theta^p \lambda^\star_x+\sum_{i=1}^M p_i l_i^\star(x, \lambda^\star_x) \right\} -  \frac{\Tilde{\mathbf{v}}(x)}{\eta}\nonumber \\
       \geq &   -\theta^p \lambda(x)+\sum_{i=1}^M p_i l_i^\star(x, \lambda(x))  -  \frac{\Tilde{\mathbf{v}}(x)}{\eta} \nonumber \\
       \geq &  -\theta^p \lambda(x)+\sum_{i=1}^M p_i l_i(x)  -  \frac{\Tilde{\mathbf{v}}(x)}{\eta} \geq \beta, \quad \forall x \in S.
\end{align}
The first inequality is true since \eqref{eq:sixthSOS} implies that $\lambda(x)\geq 0$ for $x \in S$. The second and third inequality come from \eqref{eq:SOSmiddle2} and \eqref{eq:SOSmiddle3}, respectively. Thus satisfaction of \eqref{eq:forthSOS}-\eqref{eq:sixthSOS} implies the satisfaction of \eqref{eq:decrease}. It means that $\Tilde{\mathbf{v}}$ is a DR-CBC and $\mathbf{u}$ is a associated control function. Thus for all $ x \in X_0$,
\[
\textsf{SA}_{x}^{\pi_{\mathbf{u}}}(S) \geq   \eta^{-T}\Tilde{\mathbf{v}}(x) + (\sum_{i=0}^{T-1} \eta^{-i})\beta \geq \eta^{-T}\delta + (\sum_{i=0}^{T-1} \eta^{-i})\beta, 
\]
where the first and second inequality come from \eqref{eq:approximateresult} and \eqref{eq:thirdSOS}, respectively. This completes the proof. $\hfill\square$
\end{pf}

\begin{remark}
In Theorem~\ref{thm:barrier}, we determine \( \Tilde{\mathbf{v}} \) using SOS programs under Assumption~\ref{assumption:polynomial}. 
However, for more general nonlinear systems, the program \eqref{eq:firstSOS}-\eqref{eq:seventhSOS} cannot be directly converted into an SOS program.  
Several potential approaches exist to address such cases.
One possibility is to utilize satisfiability modulo theories (SMT) solvers within a counterexample-guided inductive synthesis framework to find \( \Tilde{\mathbf{v}} \); see~\cite{barrett2010smt, de2008z3, nejati2020compositional}. Another alternative is to parameterize \( \Tilde{\mathbf{v}} \) using a neural network, enabling the synthesis of neural certificates based on constraints \eqref{eq:firstSOS}-\eqref{eq:seventhSOS}; see, for example,~\cite{mazouz2022safety, edwards2023general}. 
Note that it is non-trivial to search for DR-CBC using SMT solvers or neural networks since Eq.~\eqref{eq:decrease} involves the computation of the expectation of system evolution under a distributionally robust setting, and thus we regard it as a future direction.
\end{remark}

\section{Case Studies}\label{sec:casestudy}
In this section, we illustrate our methods with three case studies.
The first case study addresses the reach-avoid specification, employing the duality-based approach. The associated optimization problem is solved using \textsf{Gurobi}~\cite{gurobi}.
The second and third case study focuses on the safety specification, utilizing the DR-CBC approach. The SOS program involved is solved with SOSTOOLS~\cite{prajna2002introducing} and \textsf{Mosek} solver~\cite{mosek}.

\subsection{Room Temperature Control}\label{subsec:reach-avoidcase}
 \begin{table}[t]
  \centering
\caption{Statistic results for $5$ groups of parameters}\label{tab:differentpara}
  \begin{threeparttable}[b]
     \begin{tabular}{cccccc}
      \toprule
     Parameter group & $1$ & $2$ & $3$ & $4$ & $5$ \\ 
      \midrule
     ave.\ $\mathsf{RA}$ prob.\ of us (\%) & $95.7$ & $95.8$  & $95.7$ & $95.8$ & $95.8$ \\
     success rate  of us (\%) & $99$ & $100$ & $99$ & $100$ & $100$ \\
     ave.\ $\mathsf{RA}$ prob.\  of \cite{summers2010verification} (\%) & $66.2$  & $85.5$  & $88.8$  & $89.5$ & $93.0$\\
      success rate  of \cite{summers2010verification} (\%) & $54$ & $66$ & $68$ & $74$ & $86$ \\
      \bottomrule
    \end{tabular}
  \end{threeparttable}
\end{table}

 \begin{table*}[t]
  \centering
\caption{\centering{Detailed experimental results of the $2$-nd group of parameters \( N = 5 \) and \( \theta = 0.05 \) using \cite{summers2010verification}}}\label{tab:one specific}
  \begin{threeparttable}[b]
     \begin{tabular}{ccccccccccc}
      \toprule
     Round & $1$ & $2$ & $3$ & $4$ & $5$ & $6$ & $7$ & $8$ & $9$ & $10$ \\ 
      \midrule
     $\mathsf{RA}$ probability & $0.94$ & $0.96$ & $0.82$ & $0.56$ & $0.96$ & $0.96$ & $0.96$ & $0.35$ & $0.96$ & $0.35$ \\
     Distribution mean  & $0.0019$ & $-0.022$ & $0.0018$ & $0.043$ & $-0.025$  & $-0.0036$ & $-0.03$ & $0.047$ & $-0.018$ & $0.037$\\
      \bottomrule
    \end{tabular}
  \end{threeparttable}
\end{table*}

In this subsection, we consider the room temperature control problem adopted from~\cite{abate2008probabilistic} with reach-avoid specifications. Specifically, the system dynamic is given by
\[
x_{t+1}=x_t+\tau_s \left( \alpha_e (T_e-x_t) + \alpha_h (T_h-x_t) u_t \right)+w_t,
\]
where $T_e = 15$, $T_h = 50$, $\alpha_e = 8 \times 10^{-3}$, $\alpha_h = 3.6 \times 10^{-3}$ and $\tau_s = 5$. 
The initial state set,  control input set  and disturbance set are defined as
$X_0 = [23.6,23.8]$, $U=\{0,1\}$ and $W=[-0.12,0.12]$, respectively.  
 We assume that at each time step, $w_t$ follows a uniform distribution over $W$, which is unknown a priori. 
The control objective is that, within a time horizon of $T=12$, the room temperature should reach the target region $G=[24.4, 24.6]$ while staying within the safe region $S=[23,26]$ with a probability threshold $\alpha=0.9$. 

Since the distribution of the disturbance is unknown, we construct an ambiguity set based on the empirical data. Specifically, we sample the disturbance independently for $N$ times, which gives  $M$ distinct values $w_i, i = 1, \dots, M$, each associated with a frequency $p_i$. The nominal distribution is then defined as $\nu_N = \sum_{i=1}^{M} p_i \delta_{\hat{w}_i}$. In our experiment, we select five groups of parameters for the sample number $N$ and the Wasserstein ball radius $\theta$ as follows:
\[
 \{(1, 0.1), (5, 0.05), (10, 0.025), (20, 0.01), (40, 0.005)\}.
\]
The Wasserstein metric order is selected as $p = 1$.

For each group of parameters, we repeat the experiment 100 times. 
In each iteration, we solve the reach-avoid synthesis problem using both our distributionally robust approach (with respect to the Wasserstein ball defined by \( \nu_{N} \) and \( \theta \)) and the stochastic optimal control approach from~\cite{summers2010verification}, which relies on the empirical distribution \( \nu_{N} \).  
When applying each strategy, to reduce energy consumption, the controller always opts not to heat (i.e., \( u = 0 \)) whenever \( 0 \in \mathcal{U}_t(x) \) in Eq.~\eqref{eq:feasiblecontrol}. 
To evaluate the performances of the synthesized policies, 
we discretize the initial state set \( X_0 = [23.6, 23.8] \)  with a resolution of $0.01$.  
For each  initial state, we explicitly compute the reach-avoid probabilities under the actual distribution, which is unknown a priori. 
We then record the smallest reach-avoid probability across all $20$ discretized initial states. 
The experimental results are presented in Table~\ref{tab:differentpara}.

Specifically, the first and third rows of Table~\ref{tab:differentpara} show the average (across 100 repeated experiments) of the smallest reach-avoid probabilities for each parameter group using our method and the method in \cite{summers2010verification}, respectively.
The second and fourth rows record the success rate of each method. 
A synthesized policy is considered successful if the smallest reach-avoid probability exceeds the given threshold \( \alpha = 0.9 \).
Our method ensures the desired satisfaction probability in almost all trials. 
Note that the reason all parameter groups yield close satisfaction probability with our method is that every successful trial gets the optimal satisfaction probability achievable for this system under the actual uniform distribution. 
In contrast, the stochastic control framework based solely on the empirical distribution yields significantly lower average satisfaction probabilities and success rates, particularly when the sample size is insufficient.

To illustrate this more explicitly, we present the details of ten of repeated experiments for the second group of parameters (\( N = 5 \) and \( \theta = 0.05 \)) using the stochastic control approach without considering robustness. The results show that even when the sampled mean deviates only slightly from the actual distribution, the resulting controller can exhibit poor satisfaction probabilities.   
In contrast, our distributionally robust framework achieves a 100\% success rate by incorporating a small robustness radius \( \theta = 0.05 \). It is worth noting that this 100\% success rate is not guaranteed by our formal analysis, as our theoretical guarantees rely on the correctness of the ambiguity set. Nevertheless, the experimental results demonstrate that accounting for the worst-case scenarios during offline synthesis leads to consistently strong empirical performance in online applications.  
\subsection{$1$-D System Control using DR-CBC}\label{subsec:safetycase}
To illustrate the method of DR-CBC, in this subsection, we consider the one dimension system control problem for safety specifications.
The system dynamic is 
\begin{equation}
   x_{t+1}=x_t+0.1x_t^2 + u_t + w_t \nonumber
\end{equation}
where $x_t \in \mathbb{R}$ 
and $w_t, u_t \in \mathbb{R}$ are state, disturbance and control input at time instant $t$, respectively. 
The initial state set, control input set, and disturbance set are \( X_0 = [-0.5, 0] \), \( U = [0, 2] \), and \( W = [-4, 1] \), respectively. Additionally, the true but unknown distribution of \( w(t) \) is a truncated Gaussian distribution over \( W \) with a standard deviation of 2.  
The overall control objective is to ensure that, within a time horizon of \( T = 40 \), the system state is within the region \( S = \{ x \in \mathbb{R} \mid x \geq -2 \} \) with a probability exceeding the threshold \( \alpha = 0.9 \).

We take $N=5$ empirical samples from the underlying unknown distribution to define the nominal distribution 
$\nu_N$. The order of Wasserstein metric is $p=2$, and radius of the ambiguity sets are $\theta_1 = 0.1$ and $\theta_2=0.01$.
We use the SOS program \eqref{eq:firstSOS}-\eqref{eq:seventhSOS} to solve the control synthesis for safety specification.  
The degree of polynomial is selected to be $4$ and we choose $\eta=1$, $\beta = -0.0015$ and $\delta=0.96$, which ensure that 
$\delta+T\beta= 0.9$. The CBCs are computed successfully by the SOS programs under both $\theta_1$ and $\theta_2$, which can be found in \cite{chen2025distributionally}.

We simulate the system using \( \pi_\mathbf{u}^1 \) and \( \pi_\mathbf{u}^2 \) for 10000 trials, under the unknown truncated Gaussian distribution such that $\pi_\mathbf{u}^i$ is the synthesized control policy under radius $\theta_i$.
The success rates of safety specification in simulation under \( \pi_\mathbf{u}^1 \) and \( \pi_\mathbf{u}^2 \) are $99.3\%$ and $91.8\%$, respectively.  
Although our theoretical results guarantee a satisfaction probability of 0.9, this is established for all possible distributions within the ambiguity set.
In practice, our simulations achieve a higher success rate because the applied robust policy effectively accommodates the actual underlying distribution, which is not necessarily the worst-case scenario.
Moreover, when the radius is larger, more distributions are included in ambiguity set, and thus the computed control policy is more conservative.
Therefore, the success rate under \( \pi_\mathbf{u}^1 \) is higher than that under \( \pi_\mathbf{u}^2 \).
\subsection{$4$-D System Control using DR-CBC}
Finally, we consider safety control for $4$-D polynomial system in \cite{wu2025controlled} to show the scalability of proposed DR-CBC and SOS program. The system dynamic is
\begin{equation}
    \begin{aligned}
        x^1_{t+1} & = x^1_{t} + \tau_s (-x^1_{t}+(x^2_t)^3-3x^3_{t}x^4_{t}+u_t+w_t) \\
        x^2_{t+1} & =x^2_{t}+\tau_s (-x^1_{t}-(x^2_{t})^3) \\
        x^3_{t+1} & = x^3_{t} + \tau_s (x^1_{t}x^4_{t}-x^3_{t}) \\
        x^4_{t+1} & = x^4_{t} + \tau_s(x^1_{t}x^3_{t}-(x^4_{t})^3)
    \end{aligned} \nonumber
\end{equation}
where sample time $\tau_s=0.01$, $[x_t^1,x_t^2,x_t^3,x_t^4] \in \mathbb{R}^4$ 
and $w_t, u_t \in \mathbb{R}$ are state, disturbance and control input at time instant $t$, respectively. 
The initial state set, control input set, and disturbance set are \( X_0 = \{ x \in \mathbb{R}^4 \mid x_1^2+x_2^2+x_3^2+x_4^2 \leq 0.09 \} \), \( U = [-1, 1] \), and \( W = [-0.8, 0.8] \), respectively. Additionally, the true but unknown distribution of \( w(t) \) is a truncated Gaussian distribution over \( W \) with a standard deviation of $0.8$.  
The overall control objective is to ensure that, within a time horizon of \( T = 100 \), the system state remains within the region \( S =  \{ x \in \mathbb{R}^4 \mid x_1^2+x_2^2+x_3^2+x_4^2 \leq 1 \}  \) with a probability exceeding the threshold \( \alpha = 0.92 \).

We take $N=5$ empirical samples from the underlying unknown distribution to define the nominal distribution $\nu_N$.
The order of Wasserstein metric is $p=2$ and radius of the ambiguity set is $\theta = 0.1$.
We use the SOS program \eqref{eq:firstSOS}-\eqref{eq:seventhSOS} to solve the control synthesis for safety specification.  
The degree of polynomial is selected to be $4$ and we choose $\eta=1$, $\beta = -0.0004$ and $\delta=0.96$, which ensure that 
$\delta+T\beta= 0.92$.
The computed CBC can be found in \cite{chen2025distributionally}.
Then we simulate the system using the synthesized control policy \( \pi_\mathbf{u} \) for 10000 trials, under the unknown truncated Gaussian distribution.
The success rate of safety specification in simulation under \( \pi_\mathbf{u} \) is $98.13\%$, which achieves the required satisfaction probability $0.92$.

\section{Conclusion} \label{sec:con}
In this paper, we addressed the Wasserstein distributionally robust control synthesis problems for both safety and reach-avoid specifications. Our results not only established the existence of optimal policies in the distributionally robust setting but also provided computationally feasible approaches for synthesizing controllers. Experimental results validated the advantages of the distributionally robust control framework over the standard stochastic control framework, particularly in scenarios where the underlying disturbance distribution is unknown a priori and must be estimated from a limited number of samples. 

The proposed approach still has some limitations that we aim to address in future work. 
First, our framework is based on a dynamic game setting where the disturbance distribution can change dynamically. 
While effective in such scenarios, this framework results in a conservative synthesized controller when the disturbance distribution is actually time-invariant. 
Handling the time-invariant case poses additional computational challenges, as the dynamic programming principle is no longer directly applicable. 
Additionally, the proposed DR-CBC is currently limited to safety specifications, as reachability fundamentally differs from safety when deriving probabilistic lower bounds. Extending the DR-CBC framework to handle reach-avoid settings remains an important direction for future exploration.

\appendix
\section{Proof of Measurability of Stochastic Kernel}\label{appendix:measurable}
For any map of the form $(x_1,\dots,x_k)\mapsto h(x_1,\dots,x_k)$, we denote by $h(x_1,\dots, x_{m-1},\bullet,x_{m+1},\dots,x_k)$ the map $x_m\mapsto h(x_1,\dots,x_k)$. For a stochastic kernel $q(\mathrm{d}z|y)$ on $Z$ given $y$, we denote $q(\mathrm{d}z|\bullet):Y\to \mathcal{P}(Z)$ the mapping $y\mapsto q(\mathrm{d}z|y)$. The stochastic kernel is said to be measurable if $q(\mathrm{d}z|\bullet)$ is measurable. 
We first prove an intermediate result. It will be applied in proving the measurabiliy of stochastic kernel in Lemma~\ref{lemma:measurable} and the upper semi-continuity of dynamic programming value function $\mathbf{v}_t$ in Proposition~\ref{prop:uppersemi}.
\begin{mylem}\label{lem: upper semi continuous of 1_F}
        Let $Z,V$ be metrizable spaces with $V$ compact and $F \subseteq Z \times V$ a closed set. Then the real valued function
        \[\Phi: Z\times \mathcal{P}(V) \ni (z,\mu) \mapsto \int_V\mathbf{1}_F(z,v)\mathrm{d}\mu(v)\]
        is upper semi-continuous.
    \end{mylem}
\begin{pf}
Note that $Z \times \mathcal{P}(V)$ with the product topology is still metrizable.
 Let $\{(z_n,\mu_n)\}_{n \in \mathbb{N}}$ be a sequence of $\mathcal{X}=Z \times \mathcal{P}(V)$ converging to an element $(z^\star,\mu^\star)$. From Lemma 7.13 of\cite{bertsekas1996stochastic}, we can finish the proof by showing
    \[\limsup_{n\to \infty}\Phi(z_n,\mu_n) \le \Phi(z^\star,\mu^\star).\]
     Let $\varepsilon>0$ and $F_\varepsilon = \{y\in \mathcal{X}| \inf_{x\in F}d(x,y) < \varepsilon\}$. For small enough $\varepsilon$, $F_\varepsilon^c$ is a closed non-empty set. Let $g_{\varepsilon}\to [0,1]$ be the uniformly continuous real valued function on $\mathcal{X}$ that equals $1$ on $F$ and $0$ on $(F_{\varepsilon})^c$ and $g_\varepsilon(x) \in(0,1)$ for every $x \notin F\cup F_\varepsilon^c$ (the existence of $g_{\varepsilon}$ is justified by the Urysohn's lemma \cite[Lem 7.1]{bertsekas1996stochastic}). 
    For all $\varepsilon > 0$,
     \begin{align*}
       &\left|\int_Vg_{\varepsilon}(z_n,v)\mathrm{d}\mu_n(v) - \int_Vg_{\varepsilon}(z^\star,v)\mathrm{d}\mu^\star(v)\right|\le\\
            &\int_V\left|g_{\varepsilon}(u_n,v)-g_{\varepsilon}(z^\star,v)\right|\mathrm{d}\mu_n(v)+\\
            &\left|\int_Vg_{\varepsilon}(z^\star,v)\mathrm{d}\mu_n(v) - \int_Vg_{\varepsilon}(z^\star,v)\mathrm{d}\mu^\star(v)\right| \nonumber
 \end{align*}
    where both terms converge to $0$ thanks to the uniform continuity of $g_{\varepsilon}$ and the weak convergence of $\mu_n$. Thus we have proved the continuity of $(z,\mu)\mapsto \int_Vg_{\varepsilon}(z,w)\mathrm{d}\mu(w)$ for all $\varepsilon>0$. Thus we have 
    \begin{align*}
        &\limsup_n\int_V\mathbf{1}_F(z_n,v)\mathrm{d}\mu(v) \\ 
    \leq & \limsup_n\int_Vg_\varepsilon(z_n,v)\mathrm{d}\mu_n(v) 
    = \int_Vg_\varepsilon(z^\star,v)\mathrm{d}\mu^\star(v)
    \end{align*}
    where the first inequality is due to $\mathbf{1}_F\le g_\varepsilon$, the second equality is due to the continuity we have just proven.

   We have $g_\varepsilon(z^\star,\bullet)$ converges pointwisely to $\mathbf{1}_F(z^\star,\bullet)$ when $\varepsilon$ tends to $0$. As $g_\epsilon(z^\star,\bullet)$ is bounded by the constant function $1$ which is $\mu^\star$-integrable, we conclude by the dominated convergence theorem that $\int_Vg_\varepsilon(z^\star,v)\mathrm{d}\mu^\star(v)$ tends to $\int_V\mathbf{1}_F(z^\star,v)\mathrm{d}\mu^\star(v)$. Thus the real valued function $(z,\mu)\mapsto \int_V\mathbf{1}_F(z,v)\mathrm{d}\mu(v)$ is u.s.c. $\hfill\square$
    \end{pf}
Now we prove that given control policy and adversary strategy, each stochastic kernel is well-defined and thus induced a unique probability measure.
\begin{mylem}\label{lemma:measurable}
   Let $\pi=(\pi_0,\dots,\pi_{T-1}) \in \Pi$, $\gamma=(\gamma_0,\dots,$ $\gamma_{T-1}) \in \Gamma$ and $q_f:X \times U \times \mathcal{P}(W)\to \mathcal{P}(X)$ be the stochastic kernel s.t. $q_f(B|(x,u,\rho)) = \rho(f^{-1}(B)_{x,u})$ for $B \in \mathcal{B}(X)$ and $f^{-1}(B)_{x,u} = \{w\in W \mid f(x,u,w)\in B\}$.
   Then given $x_0 \in X$, there exists unique probability measure $\mathsf{Pr}_{x_0}^{\pi,\gamma} \in \mathcal{P}(\{x_0\} \times X_1 \times X_2 \cdots \times X_T)$ such that for $\underline{X}_t\subseteq X$ with $t=1,2,\dots,T$,
    \begin{equation}
        \begin{aligned}
            &\mathsf{Pr}_{x_0}^{\pi,\gamma}(x_1\in \underline{X}_1\dots x_{T} \in \underline{X}_{T}) =\\ &\int_{\underline{X}_1}\dots\int_{\underline{X}_T}q_f(\mathrm{d}x_T|(x_{T-1},\pi_{T-1}(x_{T-1}),\gamma_{T-1}(x_{T-1},\\
        &\pi_{T-1}(x_{T-1}))))\dots \times q_f(\mathrm{d}x_1|(x_0,\pi_0(x_0),\gamma_0(x_0,\pi_0(x_0)))). \nonumber
        \end{aligned}
    \end{equation}
\end{mylem}
    \begin{pf}
We first prove that the stochastic kernel $q_{\pi_t}:X \to \mathcal{P}(U)$ defined by $q(\mathrm{d}u|x) = \delta_{\pi_t(x)}$ is measurable, where $\delta_{\pi_t(x)}$ is the Dirac measure at the point $\pi_t(x)$.  Since $U$ is metrizable, the set $\{\pi_t(x)\}$ is closed for $x \in X$. Thus the measure $\delta_{\pi_t(x)}$ is well defined on $U$. Now let $\delta: U \to \mathcal{P}(U)$, $\delta(u) = \delta_u$. By Corollary 7.21.1 of \cite{bertsekas1996stochastic}, $\delta$ is continuous thus measurable. $ q_{\pi_t}=\delta \circ \pi_t$ is thus measurable as a composition of two measurable functions. 

Similarly, $q_{\gamma_t}:X\times U \ni (x,u)\mapsto \delta_{\gamma_t(x,u)} \in \mathcal{P}(\mathbb{D}) \subseteq \mathcal{P}(\mathcal{P}(W))$ is measurable. This is due to the fact that $W$ is metrizable and seperable thus $\mathcal{P}(W)$, $\mathcal{P}(\mathcal{P}(W))$, as well as $\mathcal{P}(\mathbb{D})$ are metrizable and separable thanks to \cite[prop 7.20]{bertsekas1996stochastic}, and $q_{\gamma_t}$ is the composition of $\delta:\mathbb{D}\to \mathcal{P}(\mathbb{D})$ and $\gamma_t$.

 Given $x_t,u_t = \pi_t(x_t),\rho_t = \gamma_t(x_t,\pi_t(x_t))$, the probability distribution of $x_{t+1}$ is $q_f(dx_{t+1}|(x_t,u_t,\rho_t))$. 
 For every closed set $B$ in $X$, the mapping $q_f(B|\bullet):(x,u,\gamma)\mapsto \gamma(f^{-1}(B)_{x,u})$ is u.s.c thus measurable, thanks to Lemma \ref{lem: upper semi continuous of 1_F} through replacing $Z$ by $X \times U$, $V$ by $W$ and $F$ by set $\{ (x,u,w) \mid f(x,u,w) \in B \}$. Then $q_f$ is measurable from Proposition 7.26 of \cite{bertsekas1996stochastic}. Thus stochastic kernel $x_t \mapsto q_f(\mathrm{d}x_{t+1}|(x_t,\pi_t(x_t),\gamma_t(x_t,\pi_t(x_t))))$, as a composition of $q_f,q_{\pi_t}$ and $q_{\gamma_t}$, is measurable for $t=0,1,\dots,T-1$. We conclude the lemma with proposition 7.28 of \cite{bertsekas1996stochastic}.  $\hfill\square$
    \end{pf}

\section{Proofs Omitted in Main Body}\label{appdendix:mainbodyproof}   
First, we  provide an intermediate result on proving the upper semi-continuity of dynamic programming function at each stage.
\begin{mylem}\label{lemma:upper semicontinuous int}
If $h:X \to [0,1]$ is upper semi-continuous, then the mapping $\psi:X \times U \times \mathbb{D}\to [0,1]$ such that
\[\psi:(x,u,\mu)\mapsto \int_W h(f(x,u,w))\mathrm{d}\mu(w)\]
is upper semi-continuous with respect to the product topology on $\mathcal{X}:=X \times U \times \mathbb{D}$ where $f: X\times U \times W \to X$ is the continuous system dynamic in \eqref{eq:system}.
\end{mylem}
\begin{pf}
    Consider $N\in \mathbb{N}$. For $n=1,\dots ,N$, define 
    \[
    F_n =\{ (x,u,w) \in X\times U\times W \mid h(f(x,u,w))\geq \frac{n}{N}\}.
    \]
    Then $F_n$ is closed by the definition of u.s.c function. Define $\phi_N = \frac{1}{N}\sum_{n=1}^N\mathbf{1}_{F_n}$. It is easy to check that $\phi_N(x,u,w)\le h(f(x,u,w))\le 1/N+\phi_N(x,u,w)$. Let $\{(x_n,u_n,\mu_n)\}_{n \in \mathbb{N}}$ be a sequence of $X\times U \times \mathbb{D}$ converging to $(x^\star,u^\star,\mu^\star) \in X\times U \times \mathbb{D}$. Then we have
\begin{align}
&\limsup_n\int_Wh(f(x_n,u_n,w))\mathrm{d}\mu_n(w) \nonumber \\ 
    \leq & \limsup_n\int_W\phi_N(x_n,u_n,w)\mathrm{d}\mu_n(w) +1/N \nonumber\\
    \leq & \int_W\phi_N(x^\star,u^\star,w)\mathrm{d}\mu^\star(w)+1/N \nonumber\\
    \leq & \int_Wh(f(x^\star,u^\star,w))\mathrm{d}\mu^\star(w)+1/N \label{eq:uppermiddle}
    \end{align}
    Note that the first inequality is due to $h(f(\cdot))\le 1/N+\phi_N(\cdot)$. The second is due to the fact that $\phi_N$ is a linear combination of functions of the form $\mathbf{1}_F$ and Lemma~\ref{lem: upper semi continuous of 1_F}. The last due to $\phi_N(\cdot)\le h(f(\cdot))$. As \eqref{eq:uppermiddle} holds for every $N$, let $N\to \infty$ and we complete the proof. $\hfill\square$
\end{pf}
\textbf{Proof of Proposition~\ref{prop:uppersemi}}:
\begin{pf}
Since $\mathbf{v}_T=\mathbf{1}_{G}$ and $G$ is compact, $\mathbf{v}_T$ is upper semi-continuity and  $\mathbf{v}_T(x) \in [0,1]$ for all $x \in X$. 

Assume that conditions of Proposition~\ref{prop:uppersemi} hold for $n=t+1$. For $n=t$, define function $\dot{\mathbf{v}}_t: X \times U \to \mathbb{R}$ by
\begin{equation}\label{eq:middleforoptimalcontrol}
    \dot{\mathbf{v}}_t(x,u)= \inf_{\mu \in \mathbb{D}} \mathbf{H}(x,u,\mu, \mathbf{v}_{t+1}).
\end{equation}
By inductive hypothesis, $\mathbf{v}_{t+1}: X\to [0,1]$ is upper semi-continuous. From \cite[Prop. 7.32(b)]{bertsekas1996stochastic} and Lemma~\ref{lemma:upper semicontinuous int}, we know that $\dot{\mathbf{v}}_t(x,u)$ is upper semi-continuous w.r.t. $X\times U$. 
We further define $\ddot{\mathbf{v}}_t: X \to \mathbb{R}$ by $\ddot{\mathbf{v}}_t(x) = \sup_{u \in U} \dot{\mathbf{v}}_t(x,u)$.
By \cite[Prop. 7.32(a)]{bertsekas1996stochastic} and upper semi-continuity of $\dot{\mathbf{v}}_t(x,u)$, $\ddot{\mathbf{v}}_t$ is upper semi-continuous.
From definition of integral, we have $\forall (x,u) \in X \times U$,
\[
0\leq \inf_{x' \in X} \mathbf{v}_{t+1}(x') \leq  \dot{\mathbf{v}}_t(x,u) \leq \sup_{x' \in X } \mathbf{v}_{t+1}(x') \leq 1.
\]
Thus $\ddot{\mathbf{v}}_t(x) \in [0,1]$ and $\mathbf{v}_t = \mathbf{1}_G + \mathbf{1}_{S\setminus G}\ddot{\mathbf{v}}_t \in [0,1]$  for all $x \in X$.
    Now it remains to prove that the function $\mathbf{v}_t$ is upper semi-continuous. Let $[a,+\infty]$ be an unbounded closed interval in $\overline{\mathbb{R}}$, we prove that $X_a = \mathbf{v}_t^{-1}([a,+\infty])$ is a closed set. This is true for $a > 1$ as $X_a = \emptyset$. For $a \leq 1$, let $K_a = \ddot{\mathbf{v}}_t^{-1}([a,+\infty])$ which is a closed set as $\ddot{\mathbf{v}}_t$ is upper semi-continuous. Then we have $X_a = ((S\setminus G)\cap K_a) \cup G = (S\cap K_a)\cup G$ is closed since $S$ and $G$ are closed. This completes the proof. $\hfill\square$
\end{pf}
\textbf{Proof of Proposition~\ref{prop:optimalcontrolandsuboptimaladv}}:
\begin{pf}
We first consider (1). From Proposition~\ref{prop:uppersemi} we know that function $\dot{\mathbf{v}}_t(x,u)$ in \eqref{eq:middleforoptimalcontrol} is upper semi-continuous w.r.t. $X\times U$. Then from \cite[Prop. 7.33]{bertsekas1996stochastic}, there exists $\pi^\star_t: X\to U$ which is measurable function and
\begin{equation}
    \pi^\star_t(x) \in \arg \max_{u \in U} \dot{\mathbf{v}}_t(x,u). \nonumber
\end{equation}
Thus (1) is true. Now it comes to (2). From Lemma~\ref{lemma:upper semicontinuous int}, we know that $\mathbf{H}(x,u,\mu,\mathbf{v}_{t+1})$ is upper semi-continuous w.r.t. $(x,u,\mu) \in X \times U \times \mathbb{D}$ for any $t=0,1\dots,T-1$. Since $\mathbb{D}$ is a closed-subspace of the compact space $\mathcal{P}(W)$, it's compact thus separable.
From \cite[Prop. 7.34]{bertsekas1996stochastic}, it holds that for any $\epsilon >0$, there exists a measurable function $\gamma^\star_t: X\times U \to \mathbb{D}$ such that $\forall (x,u) \in X \times U$, 
\[
\mathbf{H}(x,u,\gamma^\star_t(x,u),\mathbf{v}_{t+1}) \leq \inf_{\mu \in \mathbb{D}} \mathbf{H}(x,u,\mu,\mathbf{v}_{t+1})+\epsilon.
\]
This completes the proof. $\hfill\square$
\end{pf}

\section{The Computed Barrier Certificates and Control policies in Cases 2 and 3}
For case $2$ with $\theta_1=0.1$, the computed DR-CBC is $\Tilde{\mathbf{v}}_1(x)$ and its associated control function is $\mathbf{u}_1(x)$. Similarly, the results with $\theta_2=0.01$ are $\Tilde{\mathbf{v}}_2(x)$ and $\mathbf{u}_2(x)$.
\begin{figure*}
    \begin{align}
    \Tilde{\mathbf{v}}_1(x)&=    - 0.0002964x^4 + 0.0127x^3 - 0.1396x^2 - 0.02132x + 0.9917 \nonumber \\
      \mathbf{u}_1(x) &= 1.837e^{-6}x^4 + 3.752e^{-6}x^3 - 0.002694x^2 + 0.01221x + 1.888   \nonumber
    \end{align}
\end{figure*}
\begin{figure*}
    \begin{align}
    \Tilde{\mathbf{v}}_2(x)&=    - 1.691e^{-5}x^4 + 0.003356x^3 - 0.1835x^2 - 0.05924x1 + 0.9851 \nonumber \\
      \mathbf{u}_2(x) &= 6.812e^{-8}x^4 - 8.57e^{-6}x^3 + 0.0007303x^2 - 0.0308x + 1.517   \nonumber
    \end{align}
\end{figure*}

For case $3$, the computed DR-CBC is $\Tilde{\mathbf{v}}(x_1,x_2,x_3,x_4)$ and its associated control function is $\mathbf{u}(x_1,x_2,x_3,x_4)$.
\begin{figure*}
\begin{align}
     &\Tilde{\mathbf{v}}(x_1,x_2,x_3,x_4) =\nonumber \\
     &- 0.2498x_1^4 - 0.4521x_1^3x_2 + 6.218e^{-16}x_1^3x_3 - 2.981e^{-18}x_1^3x_4 + 0.03694x_1^3 - 0.3542x_1^2x_2^2 + 6.744e^{-16}x_1^2x_2x_3 \nonumber \\
     & + 3.973e^{-17}x_1^2x_2x_4 + 0.08168x_1^2x_2 - 0.6311x_1^2x_3^2 - 5.627e^{-5} x_1^2 x_3 x_4 - 9.227e^{-16} x_1^2 x_3 - 0.233 x_1^2 x_4^2 + 2.663e^{-18} x_1^2 x_4 \nonumber\\
     &- 0.9242 x_1^2 - 0.3872 x_1 x_2^3 + 4.514e^{-16} x_1 x_2^2 x_3 - 1.482e^{-17} x_1 x_2^2 x_4 + 0.05211 x_1 x_2^2 - 0.2759 x_1 x_2 x_3^2 + 0.0001882 x_1 x_2 x_3 x_4 \nonumber \\
     &- 7.616e^{-16} x_1 x_2 x_3 - 0.4317 x_1 x_2 x_4^2 + 4.62e^{-17} x_1 x_2 x_4 + 0.156 x_1 x_2 + 1.353e^{-15} x_1 x_3^3 - 1.112e^{-16} x_1 x_3^2 x_4 + 0.08263 x_1 x_3^2\nonumber \\
     &+ 4.98e^{-16} x_1 x_3 x_4^2 - 0.003198 x_1 x_3 x_4 + 1.376e^{-16} x_1 x_3 - 1.151e^{-17} x_1 x_4^3 + 0.001856 x_1 x_4^2 + 1.674e^{-17} x_1 x_4 - 0.0001112 x_1 \nonumber \\
     &- 0.2656 x_2^4 + 5.129e^{-16} x_2^3 x_3 + 6.286e^{-17} x_2^3 x_4 + 0.006773 x_2^3 - 0.6401 x_2^2 x_3^2 + 0.0004495 x_2^2 x_3 x_4 + 2.007e^{-16} x_2^2 x_3 \nonumber \\
     &- 0.2745 x_2^2 x_4^2 + 3.15e^{-17} x_2^2 x_4 - 0.9106 x_2^2 + 6.695e^{-16} x_2 x_3^3 + 9.216e^{-17} x_2 x_3^2 x_4 + 0.03181 x_2 x_3^2 + 5.334e^{-16} x_2 x_3 x_4^2 \nonumber \\
     & - 0.002691 x_2 x_3 x_4 + 1.647e^{-16} x_2 x_3 + 6.922e^{-17} x_2 x_4^3 + 0.003741 x_2 x_4^2 - 2.836e^{-18} x_2 x_4 - 0.0001082 x_2 - 0.5929 x_3^4 \nonumber \\
     &+ 0.001256 x_3^3 x_4 - 2.266e^{-16} x_3^3 - 0.5173 x_3^2 x_4^2 - 1.794e^{-16} x_3^2 x_4 - 0.8271 x_3^2 + 3.418e^{-5} x_3 x_4^3 - 2.969e^{-17} x_3 x_4^2  \nonumber \\
     &+ 3.805e^{-5} x_3 x_4 + 2.483e^{-18} x_3 - 0.01936 x_4^4 + 9.653e^{-18} x_4^3 - 0.9872 x_4^2 - 1.874e^{-19} x_4 + 0.9999 \nonumber \\
     &\mathbf{u}(x_1,x_2,x_3,x_4)= \nonumber \\
     &0.01773 x_1^4 - 0.0004027 x_1^3 x_2 + 3.374 e^{-16} x_1^3 x_3 - 1.196 e^{-16} x_1^3 x_4 + 0.0007027 x_1^3 - 0.0216 x_1^2 x_2^2 - 6.616 e^{-17} x_1^2 x_2 x_3 \nonumber \\
     &+ 3.953 e^{-17} x_1^2 x_2 x_4 + 0.002888 x_1^2 x_2 - 0.08656 x_1^2 x_3^2 - 3.015 e^{-5} x_1^2 x_3 x_4 - 2.731 e^{-16} x_1^2 x_3 - 0.02166 x_1^2 x_4^2 - 1.158 e^{-16} x_1^2 x_4 \nonumber\\
     &- 0.03417 x_1^2 + 0.003754 x_1 x_2^3 - 7.503 e^{-16} x_1 x_2^2 x_3 - 1.241 e^{-16} x_1 x_2^2 x_4 + 0.00351 x_1 x_2^2 - 0.002516 x_1 x_2 x_3^2 \nonumber \\
     &+ 0.0009659 x_1 x_2 x_3 x_4 - 3.586 e^{-18} x_1 x_2 x_3 - 0.0003268 x_1 x_2 x_4^2 - 1.388 e^{-16} x_1 x_2 x_4 - 0.005415 x_1 x_2 - 2.386 e^{-16} x_1 x_3^3 \nonumber \\
     &- 1.456 e^{-16} x_1 x_3^2 x_4 + 0.005017 x_1 x_3^2 - 7.791 e^{-16} x_1 x_3 x_4^2 - 0.006335 x_1 x_3 x_4 - 1.291 e^{-15} x_1 x_3 - 1.341 e^{-16} x_1 x_4^3 \nonumber \\
     &+ 0.002866 x_1 x_4^2 + 1.0 e^{-16} x_1 x_4 + 0.001509 x_1 + 0.04643 x_2^4 + 2.58 e^{-17} x_2^3 x_3 - 2.009 e^{-16} x_2^3 x_4 + 0.0003539 x_2^3 - 0.02193 x_2^2 x_3^2 \nonumber \\
     &+ 0.000234 x_2^2 x_3 x_4 - 2.226 e^{-17} x_2^2 x_3 + 0.05167 x_2^2 x_4^2 - 2.946 e^{-17} x_2^2 x_4 - 0.07807 x_2^2 + 5.043 e^{-17} x_2 x_3^3 + 6.089 e^{-17} x_2 x_3^2 x_4 \nonumber \\
     &+ 0.0002895 x_2 x_3^2 - 1.03 e^{-16} x_2 x_3 x_4^2 - 0.001483 x_2 x_3 x_4 - 3.494 e^{-17} x_2 x_3 - 3.811 e^{-16} x_2 x_4^3 + 0.0006056 x_2 x_4^2 + 2.91 e^{-16} x_2 x_4 \nonumber \\
     &- 0.0001431 x_2 + 0.01818 x_3^4 + 0.0003328 x_3^3 x_4 - 6.936 e^{-17} x_3^3 - 0.02228 x_3^2 x_4^2 + 9.971 e^{-17} x_3^2 x_4 - 0.03461 x_3^2 - 0.0001048 x_3 x_4^3 \nonumber \\
     &+ 2.35 e^{-17} x_3 x_4^2 - 1.252 e^{-5} x_3 x_4 + 9.886 e^{-17} x_3 + 0.04612 x_4^4 + 1.458 e^{-18} x_4^3 - 0.07739 x_4^2 - 1.024 e^{-17} x_4 + 0.9907 \nonumber
\end{align}
\end{figure*}
\bibliographystyle{plain}
\bibliography{main}

\end{document}